\documentclass[preprint]{aastex}

\usepackage{subfigure}

\shorttitle{The Search for Celestial Positronium}
\shortauthors{Ellis & Bland-Hawthorn}

\newcommand\lsim{\mathrel{\hbox{\rlap{\hbox{\lower4pt\hbox{$\sim$}}}\hbox{$<$}}}}
\newcommand\pos{$e^{+}$}
\newcommand\el{$e^{-}$}
\newcommand\psla{Ps Ly$\alpha$}
\newcommand\psa{Ps$\alpha$}
\newcommand\gsim{\mathrel{\hbox{\rlap{\hbox{\lower4pt\hbox{$\sim$}}}\hbox{$>$}}}}

\begin{document}

\title{The Search for Celestial Positronium via the Recombination Spectrum} 

\author{S.~C. Ellis and J. Bland-Hawthorn}
\affil{Sydney Institute for Astronomy, School of Physics, The University of Sydney, NSW 2006, Australia\\sce@physics.usyd.edu.au,  jbh@physics.usyd.edu.au}

\begin{abstract}
Positronium is the short-lived atom consisting of a bound electron-positron pair.  In the triplet state, when the spins of both particles are parallel, radiative recombination lines will be emitted prior to annihilation.  The existence of celestial positronium is revealed through gamma-ray observations of its annihilation products. These observations however have intrinsically low angular resolution.
In this paper we examine the prospects for detecting the positronium recombination spectrum.  Such observations have the potential to reveal discrete sources of \pos\ for the first time and will allow the acuity of optical telescopes and instrumentation to be applied to observations of high energy phenomena.

We review the theory of the positronium recombination spectrum and provide formul\ae\ to calculate expected line strengths from the \pos\ production rate and for different conditions in the interstellar medium.  We estimate the positronium emission line strengths for several classes of Galactic and extragalactic sources.  These are compared to current observational limits and to current and future sensitivities of optical and infrared instrumentation.
We find that observations of the \psa\ line should soon be possible due to recent advances in near-infrared spectroscopy.  
\end{abstract}


\section{INTRODUCTION}

Positronium (Ps) is the short-lived atom consisting of a bound electron-positron pair.  Its existence was predicted by \citet{moh34} shortly after the discovery of the positron by \citet{and33}.  Furthermore, \citet{moh34} suggested that Ps should be observable by its recombination spectrum and discussed its properties.  Ps was first observed by \citet{deu51} in a laboratory experiment designed to measure the lifetimes of positrons in gases.

Ps is unstable and annihilates into two or more photons with a combined energy of 1022 keV, i.e. the rest mass of the atom (see \S~\ref{sec:ann}).  However, as discussed in \S~\ref{sec:phys}, under certain conditions Ps is stable for long enough that electronic transitions can occur between the quantum states of the system, giving rise to a recombination spectrum.  Since the reduced mass of Ps is half that of H, the wavelengths are twice those of the corresponding H series.  For example, Ps Lyman $\alpha$ (\psla) has a wavelength of 2431\AA\ and Ps Balmer $\alpha$ (\psa) has a wavelength of 1.313$\mu$m. 

\psla\ was first observed in laboratory experiments by \citet{can75}.  So far, however, no detections have been made of recombination lines from celestial Ps.  Nevertheless the existence of celestial Ps is inferred from observations of the 511keV annihilation line and continuum annihilation radiation seen both from the Sun (\citealt{chu73}) and from the Galactic centre (\citealt{lev78}), combined with the fact that in most astrophysical environments \pos\ are expected to form Ps before annihilation (cf.\ \S~\ref{sec:phys}).  The successful observation of Ps recombination lines would not only be of interest in its own right, but would also have important consequences for many areas of astrophysics.  We now describe the motivation behind a search for Ps recombination lines.

 Current knowledge of naturally occurring antimatter from astrophysical sources is limited to the poor spatial resolution ($\sim 3^{\circ}$) of $\gamma$-ray satellites.  The formation of Ps, however, provides a natural heterodyne process that converts the $\gamma$-ray energies associated with the production of \el-\pos\ pairs, to the optical/ near-infrared energies of Ps recombination. Thus observation of Ps recombination lines would allow the superior spatial and spectral resolution of optical/ near-infrared technologies to be applied to high energy astrophysical phenomena.

As a specific example, Ps observations could provide an important physical understanding of jets.  It is still uncertain whether high energy radio jets consist of an \el-p plasma or \el-\pos\ or some combination of both.  Most attempts to answer this question have had to rely on indirect methods of searching for the presence of \pos, such as arguments based on the total energy budget of the jets (e.g.\ \citealt{cel93}; \citealt{rey96}; \citealt{war98}; \citealt{hir05}), with inconclusive results.  Direct searches for the 511keV annihilation signature  in jets (\citealt{mar07}) have likewise failed to provide conclusive results due to the poor point source sensitivity of $\gamma$-ray telescopes.  If the positrons from the jet thermalise when the jet impacts the surrounding medium allowing the formation of Ps, and if the acuity of optical observations can be brought to bear on this problem to allow a successful detection of Ps recombination lines, this would have significant implications for our understanding of the physical processes responsible for the production of jets -- processes which take place deep within the AGN (\citealt{bla95}).

 The origin of the positrons responsible for the annihilation radiation observed from the Galactic centre is uncertain.  The current state-of-the-art observations are from the ESA's INTEGRAL $\gamma$-ray observatory, which has detected extended Ps annihilation radiation centred on the Galactic centre with a FWHM of $\approx 8^{\circ}$ (\citealt{kno05}; \citealt{wei08b}).    The results are most consistent with a source of \pos\ originating from old stellar populations such as type Ia supernov\ae , classical nov\ae , and low mass X-ray binaries, although more exotic sources, such as dark matter annihilation, cannot be ruled out.  The angular resolution of INTEGRAL SPI is $\sim 3^{\circ}$, and is therefore rather insensitive to the detection of point sources.   The detection of Ps recombination lines would allow discrete point sources of \pos\ to be searched for on sub-arcsecond scales (see \S~\ref{sec:sources}) allowing direct  confirmation of the \pos\ production of specific sources for the first time.


Despite the
 benefits of detecting Ps recombination lines there have been very few attempts to do so, and there has been little work on the expected astrophysical signatures.   \citet{mcc84} reviews the possibility of the optical detection of Ps, in particular the \psla\ line.  The conclusion reached is that a visual extinction of 1 or 2 magnitudes is enough to make the Balmer and Paschen lines more intense than the intrinsically more luminous Lyman lines.  Since most previous studies have focussed on the Galactic centre as  the most promising source of Ps recombination line emission (so far the only source other than the Sun in which Ps annihilation has been observed) and that at the Galactic centre the visual extinction is very high, it is not surprising that this idea has remained largely forgotten.  \citet{burd97} raised the possibility of optical/ IR detection of Ps again, and made predictions of the expected fluxes of the most important lines.  However, the only published attempt to detect Ps recombination lines was made by \citet{pux96} who searched for Ps Paschen-$\beta$ from the Galactic centre in the $K$ band, and whose non-detection provides the only direct  experimental constraints on the emission.

Until recently, the prospect of a near-infrared (NIR) detection of the Balmer series of Ps was rather bleak due to the relative insensitivity of  NIR detectors and the strong foreground emission from the atmosphere.  However, we argue that the time is now right for a reappraisal of this important window into the high energy universe.  This renewed interest is motivated by several recent and imminent advances in technology.  NIR detectors have improved remarkably and are now capable of dark currents as low as 1\el\ per 1000 s, and read noises of $\approx 8$\el (\citealt{smi06}).  Adaptive optics at NIR wavelengths can now deliver high Strehl ratios improving the signal to noise for faint sources.  The future generation of ELTs will deliver AO corrected images of fields of view of tens of arcminutes with huge gains in sensitivity from the vast collecting areas.  The JWST will provide unprecedented sensitivity at NIR wavelengths (\citealt{gar06}).  Recent breakthroughs in photonic technologies promise superb background suppression in the NIR in the near future (\citealt{bland04}; \citealt{ell08}).   
  
The purpose of this paper is to assess the probability of detecting Ps recombination spectra in light of these advances and to identify the most fruitful strategies  and targets for an observational campaign.  We collect together sources from a broad literature published over many decades and combine these to calculate the expected astrophysical signatures of Ps. The next section reviews the pertinent physical properties of Ps.  We then discuss possible sources of Ps recombination line emission, both Galactic and extragalactic, including predictions on the strengths of these sources.  Section~\ref{sec:expt} then assesses the prospects of detection and observational strategies.  Finally we summarise our findings in section~\ref{sec:summ}.

\section{PHYSICS OF THE POSITRONIUM RECOMBINATION SPECTRUM}
\label{sec:phys}

\subsection{Formation of Ps}

The fate of positrons in astrophysical environments  depends on their energy and the ionisation fraction of the surrounding medium.    Positrons are formed in highly energetic processes ($\ge 1022$ keV as part of an \el-\pos\ pair), and initially may be highly relativistic.  In order that Ps may be formed the \pos\ must first lose energy.  We consider two stages in the formation of Ps, the in-flight phase during which the positrons are losing energy, and the thermalised phase, during which the \pos\ are in thermal equilibrium with the surrounding medium.

\subsubsection{In-flight phase}

Positrons can lose energy via collisions with electrons or by excitation and ionisation of atoms and molecules.  During this energy loss phase, once the \pos\ has an energy of less than a few tens of eV, Ps can be formed via charge transfer with H, H$_{2}$ or He.  \citet{gue05} and \citet{bus79} show that the fraction of \pos\ forming Ps in flight is a strong function of the ionisation fraction of the medium.  In a completely neutral medium consisting of atomic H, approximately 95 per cent of \pos\ form Ps (via charge exchange).  If the same medium were 50 per cent ionised, less than 10 per cent of \pos\ would form Ps (\citealt{gue05}).   If the medium is collisionally ionised, then the ionisation fraction depends on the temperature and density of the gas.  However, \citet{mcc84} notes that in most cases of \pos\ generation from an astrophysical source (e.g.\ supernov\ae, microquasars, etc.), a radiative source capable of photoionising the surrounding medium will be involved.    In this case the ionisation fraction of the medium close to the source is of order unity and independent of gas temperature.  Consequently, almost all \pos\ will survive the in-flight phase and will eventually thermalise with the surrounding medium.

\subsubsection{Thermalised phase}

Once thermalised, \pos\ may form Ps via radiative recombination or via charge exchange.  The relative efficiencies of these processes are highly dependent on temperature (\citealt{bus79}; \citealt{mcc84}; \citealt{gue05}).  For a collisionally ionised gas, Ps formation by radiative recombination with free electrons dominates charge exchange at $T\lsim 10^{4}$ K (\citealt{gue05}).  However, for a photoionised gas radiative recombination is dominant at all temperatures $<10^{6}$ K (\citealt{mcc84}).  

It is possible for \pos\ to annihilate directly with either free or bound \el\ without first forming Ps.  Direct annihilation with free \el\ has a cross section that is an order of magnitude lower than radiative recombination for $T<10^{5}$ K (\citealt{gue05}).  Direct annihilation with bound \el\ can only take place at low temperatures ($\lsim 10^{3}$ K; \citealt{gue05}).  Therefore for most astrophysical environments ($\sim 10^{3}$--$10^{6}$ K) Ps formation will be the dominant process leading to annihilation.  


\subsubsection{Formation in dust grains}

Ps can form when a \pos\ is captured by a dust grain.  Dust grains have a much larger cross-section than the competing effects of recombination or charge-exchange and hence can play a significant role in the formation of Ps.  The effects of Ps formation in dust was first discussed by \citet{zur85} and the most recent treatment is given by \citet{gue05}.  We summarise the relevant findings here.

\citet{gue05} give four possible reactions between \pos\ and dust grains, \pos\ re-emission, \pos\ backscattering, Ps formation in the grain, Ps ejection.  Ps formed in the grain always annihilates  via para-Ps (see \S~\ref{sec:ann}) since ortho-Ps will quickly convert to para-Ps via ``electron pick-off'' of the surrounding electrons in the dust grain.  Ejected Ps will annihilate in the usual 3:1 ratio of ortho:para-Ps (see \S~\ref{sec:ann}).  Thus only the ejected Ps will contribute to any optical/NIR emission.

The significance of annihilation via Ps formation in dust grains is dependent on the temperature of the ISM. \citet{gue05} find that the role of dust is insignificant at low ISM temperatures, since in this case dust grains are thought to be positively charged, which reduces the chance of \pos\ capture.  In the warm ionised ($\sim8000$K) and hot ($\sim10^{6}$K) phases of the ISM dust is thought to be negatively charged and \pos\ capture can become significant.   The reaction rate for radiative recombination, direct annihilation and capture by dust grains are given by \citet{gue05} as $1.2 \times 10^{-12}$, $1.7\times10^{-13}$ and $4.6 \times 10^{-14}$ cm$^{3}$ s$^{-1}$ respectively for the warm ionised medium, and $1.3 \times 10^{-14}$, $1.8\times 10^{-14}$ and $2.4 \times 10^{-13}$ cm$^{3}$ s$^{-1}$ for the hot medium.  Thus in the warm phases recombination is dominant but in the hot phases dust process are the dominant form of Ps formation.  

We also wish to know the relative rates of the reactions that will generate recombination lines, and this requires annihilation via ortho-Ps (\S~\ref{sec:ann}).  \citet{gue05} estimate that about $\sim 10$ per cent of Ps formed in dust will be ejected before annihilation.  Taking this factor and the 3:1 ortho:para ratio into account gives reaction rates in the hot phase of the ISM of $9.8 \times 10^{-15}$ and $1.8 \times 10^{-14}$ cm$^{3}$ s$^{-1}$ for the formation of ortho-Ps via radiative recombination and dust capture respectively.

\subsection{Lifetime and Annihilation of Ps}
\label{sec:ann}

Ps is unstable and will eventually annihilate.  The process of annihilation is governed by the principle of invariance under charge conjugation (\citealt{yan50}; \citealt{wol52}; \citealt{berk80}) and the mechansim depends on the total spin of the atom (see e.g. \citealt{deb54}).  Para-Ps annihilates into an even number of photons, with the most likely decay into two photons, whereas ortho-Ps annihilates into an odd number of photons with the most likely decay being into three photons.  Furthermore conservation of energy and momentum requires that para-Ps decays into two photons of equal energy, giving rise to the 511keV emission line, 
whereas the energies of the three ortho-Ps  annihilation photons can be any three energies that add to 1022keV, giving rise to a continuum of $\gamma$-ray photons $<511$keV.  
The different decay mechanisms give rise to different lifetimes for para- and ortho-Ps.  Para-Ps in the ground state has a lifetime of $1.26\times10^{-10}$s and ortho-Ps a lifetime of $1.41 \times 10^{-7}$s (e.g.\ \citealt{berk80}).
The lifetimes of excited states are longer by a factor $n^{3}$ (\citealt{whe46}).
Since para-Ps is a singlet state and ortho-Ps is a triplet state the relative weights of ortho-Ps to para-Ps are 3:1.

\subsection{Ps Recombination Spectrum}
\label{sec:spec}

Ps has a reduced mass of $m_{\rm e}/2$, so the energy levels of Ps are almost exactly half those of hydrogen, which has a reduced mass $\approx m_{\rm e}$.  Correspondingly, the wavelengths of the recombination spectrum are twice those of hydrogen.  This has been confirmed through laboratory observations of the \psla\ line at 2431\AA\ by \citet{can75}.


Similarly, the radiative lifetimes are twice those of hydrogen. We have calculated the radiative lifetimes of Ps using the atomic hydrogen transition probabilities listed by \citet{wie66}.  We use the `average' values tables which are most applicable to astrophysical situations if it is assumed that all atomic sub-states are occupied according their statistical weights.  In Table~\ref{tab:lifetimes} we show the radiative and annihilation lifetimes for states $n=1$--6 and $l=0$ (the annihilation lifetimes are almost entirely independent of $l$), and the ratios of these quantities.  The ratio of annihilation lifetime to radiative lifetime ($T/\tau$) gives the approximate number of photons that \emph{could} be emitted from that state before decay.  For triplet Ps all states could emit more than one photon (significantly more for the lower energy states) before decay.  Thus most triplet Ps annihilates from its ground state.  Singlet Ps would decay before any radiation is emitted.  We note however that in most cases Ps will form via case-A recombination and therefore after radiative recombination down to the $n=1$ level (perhaps via multiple transitions), we do not expect any re-excitation followed by further re-radiation, since this would require the presence of other Ps atoms nearby, which is unlikely in most cases.  We also note that the probability of spin-flip before radiative transition is very small\footnote{see Burdyuzha, Durouchoux \& Kauts 1999, http://jp.arxiv.org/abs/astro-ph/9912550}; the Einstein $A$ coefficient for spontaneous spin-flip, $A=3.3\times 10^{-8}$ s$^{-1}$.   Collisionally induced spin-flip is highly unlikely as the lifetime of Ps is so short (\S~\ref{sec:ann}), and would require densities of particles in the ISM far in excess of those observed.  Therefore we do not expect any spin-flip to occur before annihilation.

\begin{table}
\center
\caption{\label{tab:lifetimes}The radiative and annihilation lifetimes of Ps for various energy levels.}
\begin{tabular}{llllll}
& & \multicolumn{2}{c}{Singlet Ps} & \multicolumn{2}{c}{Triplet Ps} \\
& Radiative lifetime & Annihilation lifetime & Ratio & Annihilation lifetime & Ratio \\
$n$ & $\tau$/s & $T$/s & $T/\tau$ &  $T$/s & $T/\tau$ \\ \hline
 1 & -- &                                    $1.25 \times 10^{-10}$  & --          & $1.41 \times 10^{-7}$ & -- \\
 2 & $4.26 \times 10^{-9}$ & $10.0 \times 10^{-10}$ & 0.23     & $1.1 \times 10^{-6}$ & 265 \\
 3 & $8.12 \times 10^{-8}$ & $3.37 \times 10^{-9}$   & 0.041   & $3.8 \times 10^{-6}$ & 46.9 \\
 4 & $6.17 \times 10^{-7}$ & $8.00 \times 10^{-9}$   & 0.013   & $9.0 \times 10^{-6}$ & 14.6 \\
 5 & $2.90 \times 10^{-6}$ &   $1.56 \times 10^{-8}$   & 0.0053 & $1.76 \times 10^{-5}$ & 6.03 \\
 6 & $1.04 \times 10^{-5}$ & $2.70 \times 10^{-8}$   & 0.0026 & $3.05 \times 10^{-5}$ & 2.93 
\end{tabular}
\end{table}

\citet{wal96} calculate line widths and intensities for the radiative recombination spectrum of Ps in the limit of thermal positrons and low densities, including annihilation from excited states.  The lines strengths are assumed to be dominated by electron-positron radiative recombination for free electrons, with a small fraction of Ly-$\alpha$ emission resulting from electron capture through interaction of a positron with H or He.

For a given source producing $r$ positrons per second, the flux of Ps radiative emission in photons s$^{-1}$ m$^{-2}$ $\mu$m$^{-1}$ would be,
\begin{equation}
\label{eqn:flux_source}
f_{\lambda}=r f_{\rm Ps} \frac{3}{4} \frac{\alpha}{\beta} \times \frac{1}{4 \pi D_{\rm L}^2}\times \frac{1}{2.512^{A}}\times \frac{1}{\Delta\lambda},
\end{equation}
where $f_{\rm Ps}$ is the fraction of \pos\ which form Ps after thermalisation, the factor 3/4 is the fraction of Ps in the triplet state, $\alpha$ is the number of emitted Ps Ly$\alpha$ photons for each Ps atom, $\beta$ is the ratio of the intensity of Ps Ly$\alpha$ to the line in question, $D_{\rm L}$ is the luminosity distance of the source, the factor $1/2.512^{A}$ accounts for an absorption coefficient of $A$, and $\Delta\lambda$ is the line width in microns.

We note an important point:  $\gamma$-ray observations of the 511keV line are observing the annihilation of para-Ps, whereas observations of the radiative recombination spectrum are observing ortho-Ps.  Since the relative weights of para-Ps to ortho-Ps are 1:3 (singlet vs. triplet states), but there are 2$\times$511keV photons for each annihilation, but only one radiative recombination photon,
the recombination line flux ($f_{\lambda}$) can be related to the $\gamma$-ray line flux ($f_{511}$ photons s$^{-1}$ m$^{-2}$) by,
\begin{equation}
\label{eqn:wallyn}
f_{\lambda}=f_{511} \times \frac{3}{2} \times \frac{\alpha}{\beta} \times \frac{1}{2.512^{A}} \times \frac{1}{\Delta\lambda},
\end{equation}
The line width is given by \citet{wal96} as 
\begin{equation}
\label{eqn:linewidth}
\Delta\lambda=\lambda \times 7.8 \times 10^{-4} \left(\frac{T}{10^{4} {\rm K}}\right)^{0.44},
\end{equation}
which is somewhat narrower than the assumption of purely thermally broadened Ps emission, and results from Monte-Carlo simulations of the combination process by \citet{gue91} which show that the electron-positron system is not in exact thermal equilibrium with the gas.

\citet{wal96} tabulate values of $f_{\rm Ps}$,  $\alpha$ and $\beta$ for different temperatures, which we include here for convenience in Table~\ref{tab:temp}  along with the line widths of \psla\ and \psa\ calculated from equation~\ref{eqn:linewidth}.  The positronium fraction, $f_{\rm Ps}$, is calculated from the reaction rates for radiative recombination, charge exchange with H and He  and direct annihilation given by \citet{wal96}.  The fraction of Ps which undergoes a Ly$\alpha$ transition, $\alpha$, is given by \citet{wal96} in their table~5.  The ratio of the intensity of \psla\ to \psa, $\beta$, is derived from the absolute intensities of the transitions listed in table~6 of \citet{wal96}.  Note that there are two values of $f_{\rm Ps}$ given, one for a neutral medium which includes Ps formation via charge exchange, and one for an ionised medium in which Ps only forms by radiative recombination.

\begin{table}
\center
\caption{A summary of the relevant parameters for calculating the recombination line fluxes for different temperatures, extracted from \protect\citet{wal96}.  The fraction of \pos\ forming Ps in an ionised medium; a neutral medium; the Ps Ly$\alpha$ fraction, $\alpha$; the ratio of line intensities I(Ly-$\alpha$)/I(\psa), $\beta$; \psla\ and \psa\ line widths.}
\label{tab:temp}
\begin{tabular}{lcccccc}
$T/$ K & $f_{\rm Ps}$ ionised &$f_{\rm Ps}$  neutral &$\alpha$ & $\beta$ &$\Delta\lambda$(\psla)/ \AA & $\Delta\lambda$(\psa)/ \AA \\ \hline
$1\times10^{3}$   &0.41&0.41& 0.45 & 1.816 & 0.69 &3.72 \\
$5\times10^{3}$ &0.87&0.88& 0.37 & 2.192 & 1.40 & 7.55 \\
$1\times10^{4}$   &0.86&0.99& 0.32 & 2.391 & 1.90 & 10.234 \\
$2\times10^{4}$   & 0.84&1.0&0.28 & 2.586 & 2.57 & 13.89 \\
$1\times10^{5}$   & 0.75&-&0.18 & 2.731 & 5.22& 28.20 \\
$1\times10^{6}$   & 0.41&-&0.09 & 2.242 & 14.38 & 77.67 \\
\end{tabular}
\end{table}

For \psla\ and \psa\ we assume $A_{2433}=1.716 A_{\rm V}$ and $A_{\rm J} = 0.282 A_{\rm V}$ respectively (\citealt{rie85}).

%
%

\section{SOURCES OF POSITRONIUM}
\label{sec:sources}

We now consider the expected Ps recombination line emission from various astrophysical sources.  We discuss both Galactic and extragalactic sources and calculate fluxes for \psla\ and \psa\ using the formul\ae\ given in section~\ref{sec:spec}.  We quote our results in terms of line flux densities, i.e.\ the total line strength divided by the line width.  This allows for a more meaningful comparison to the source continuum spectrum and the atmospheric background, both of which are found to dominate the Ps emission line fluxes.

\subsection{Galactic sources}
\label{sec:galactic}

So far the only detections of Ps outside laboratory experiments are from the Sun (first observed by \citealt{chu73}) and the Galactic centre (first identified by \citealt{lev78}).   All detections have been via annihilation radiation; no detection of the optical or NIR recombination lines exists.

Since the early pioneering detections of 511keV radiation much progress has been made, with significant increases in precision and sensitivity (see for example \citealt{mur05} for recent observations of 511keV emission from the Sun).  For the the Galactic emission  the current state-of-the-art observations are from the ESA's INTEGRAL $\gamma$-ray observatory, which has detected extended Ps 511keV line emission and continuum emission centred on the Galactic centre (\citealt{wei08b}).   Broadly, the distribution of 511keV radiation toward the Galactic centre is extended with
 a FWHM of $\approx 8$ degrees (\citealt{kno05}) and the continuum emission is consistent with this (\citealt{wei06}).  More detailed analyses reveal a bulge and a disc component to the annihilation radiation (see \citealt{wei08b} for a summary).    However, the angular resolution of INTEGRAL SPI is $\approx 3$ degrees, and is therefore insensitive to the detection of point sources.  
The results are most consistent with a source of positrons originating from old stellar populations such as type Ia supernov\ae , classical nov\ae , and low mass X-ray binaries, although more exotic sources cannot be ruled out.  It is suggested that the reason for the diffuse nature of the 
annihilation radiation is that e$^{+}$ are long lived, and the reaction rates with the ISM are small (\citealt{gue05}) and positrons can diffuse long distances from their sources before thermalisation (\citealt{jea06,gil07}).  Thus e$^{+}$ have a large mean free path before interaction and annihilation.
Even allowing for the longevity of \pos\ in the interstellar medium, and the resulting diffuse nature of Ps annihilation radiation, it is likely that some fraction of the Ps annihilation radiation results from
 discrete sources.   (i) It is expected that individual energetic sources will emit \pos\ and therefore these should be brighter than any diffuse background.  (ii) The Sun is known to emit Ps 511keV radiation, and therefore more energetic sources of \pos\ will also likely emit Ps 511keV radiation.  (iii) There is a correlation between an asymmetry in the 511keV radiation originating from the inner Galactic disc and the distribution of $\gamma$-ray selected LMXBs (\citealt{wei08}).  (iv)
The ISM is very inhomogeneous; the range of densities ranges from approximately $10^{3}$ electrons m$^{-3}$ in the hot diffuse gas ($\approx 10^{6}$K) to $2 \times 10^{7}$ atoms m$^{-3}$ in cold gas ($\approx 100$K), and even higher in molecular clouds.  Similarly the sizes of these regions vary significantly, from $\sim 1$pc to $\sim 100$pc, and we anticipate that some clumps will be more favourable 
for Ps formation, and therefore more luminous,  than others.  (v) The history of both the
 X-ray and the IR background cautions us to suspect that a large number of faint point sources may masquerade as a diffuse background (\citealt{sha91,hau01}). 

\citet{kno05} have searched for point sources within the INTEGRAL data.  They find that 8 individual point sources could describe the emission seen in the Galactic bulge, with a combined flux of $f_{511} = 11$ photons s$^{-1}$ m$^{-2}$ , or roughly $1.4$ photons s$^{-1}$ m$^{-2}$ per source.
In addition to this blind survey, \citet{kno05} also examine candidate sources including Sgr A*, microquasars, LMXBs, pulsars, SNRs, galaxies and AGN.  No point sources are detected down to typical limits of $\approx 0.7$--$3.8$ photons s$^{-1}$ m$^{-2}$.  The upper limit for Sgr A* is
 1 photons s$^{-1}$ m$^{-2}$.  
These results suggest that if point sources are responsible for the observed diffuse emission they are likely to be more than 8 and fainter than $\approx  1$ photons s$^{-1}$ m$^{-2}$.  Assuming all sources to be of equal strength we therefore empirically relate the 511keV flux of individual point sources to the total number of sources, $N$, by,
\begin{equation}
\label{eqn:fluxno}
f_{511}\approx\frac{11.2}{N} {\rm  \ photons\ s^{-1}\ m^{-2}.}
\end{equation}

\begin{figure}
\begin{center}
\subfigure[\psla]{
\centering \includegraphics{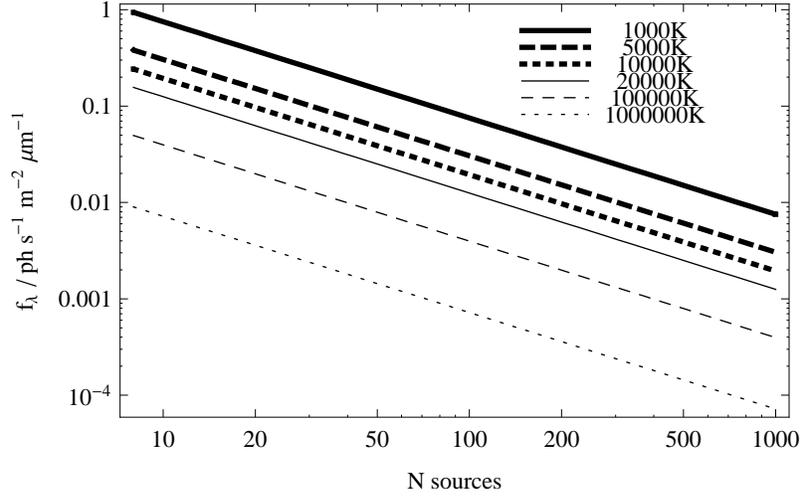}
}\\
\subfigure[\psa]{
\centering \includegraphics{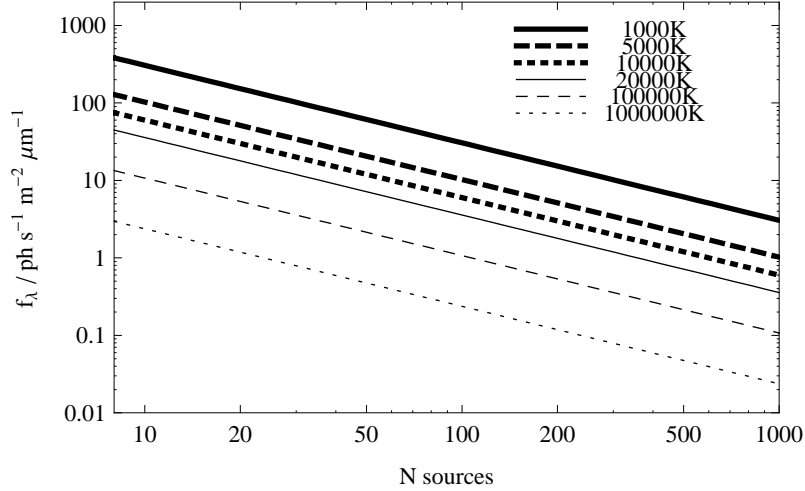}
}
\caption{The line flux density of \psla\ and \psa\ assuming that all the Galactic 511keV radiation is due to $N$ equal point sources, and that the point sources are obscured with $A_{\rm V} = 5$mag at the Galactic centre.}
\label{fig:pointsources}
\end{center}
\end{figure}

Combining equation~\ref{eqn:fluxno} with equations~\ref{eqn:wallyn} and \ref{eqn:linewidth} and using the parameters listed in Table~\ref{tab:temp} we can calculate the line flux per source as a function of the number of sources.  This is plotted in Figure~\ref{fig:pointsources} for both \psla\ and \psa\ for a range of temperatures.  It was assumed that the extinction towards the Galactic centre was $A_{\rm V}=5$ mag, although we note that there are several windows of low extinction (\citealt{dut02}), such as Baade's window, which may be used to obtain deeper observations of the Galactic centre.  It is clear that \psa\ emission is much brighter than \psla\, despite the fact that \psla\ is intrinsically brighter.  For a temperature of $10^{4}$ K the flux density is higher for \psa\ for any visual extinction $A_{\rm V} > 1.2$ mag, for the total integrated line strength \psa\ is brighter for $A_{\rm V} >0.66$ mag.  

We defer an assessment of the feasibility of observing such sources to section~\ref{sec:expt}, but we note as a guide sources that with flux densities $ f_{\lambda} \gg 1$ ph s$^{-1}$ m$^{-2}$ $\mu$m$^{-1}$ may be considered bright, whereas those with $ f_{\lambda} \ll 1$ ph s$^{-1}$ m$^{-2}$ $\mu$m$^{-1}$ may be considered faint, c.f\ \S~\ref{sec:expected}.  First we discuss individual types of sources and make estimates of their likely Ps recombination line emission.   \citet{gue05} list the following processes for the production of \pos\,a) $\beta^{+}$ decay of radioactive nuclei; b) $\pi^{+}$ decay into $\mu^{+}$, which decays and gives off a positron; c) pair (electron-positron) production through photon-photon interactions; d) pair production by the interaction of an electron with a strong magnetic field.  All these processes are thought to occur naturally in the Universe.  In the Galaxy, the main sources of \pos\ are thought to be supernov\ae , nov\ae , microquasars and pulsars, cosmic rays and $\gamma$-ray bursts, in that order (\citealt{gue05}).  To this list we add the decay of $^{26}$Al, which is produced via nucleosynthesis in core-collapse supernov\ae, nov\ae, asymptotic giant branch stars and massive stars (\citealt{kno99}), and is expected to make a significant contribution to the \pos\ content of the Galactic disc (\citealt{kno05}).
We also add decaying super-symmetric dark matter particles (\citealt{boe04}; \citealt{bea05}).  We discuss each of these in turn below, except for cosmic rays and dark matter decays.  Both these processes are intrinsically diffuse, and thus while possibly significant in terms of the Galactic positron budget they are unlikely candidates for Ps emission line observations.

Most of the sources listed above inject \pos\ into the ISM, where they thermalise before forming Ps and annihilating.  Once in the ISM the \pos\ will diffuse along magnetic field lines.  The diffusion depends primarily on the conditions in the ISM (assuming a restricted energy range for the injected \pos\ $\sim 1$ MeV) and not on the source of \pos, thus the average distance over which the \pos\ will be essentially identical for all injecting sources.  We therefore preface our discussion of the individual sources with a discussion of the diffusion of \pos\ in the ISM.

\subsubsection{Diffusion of \pos\ in the interstellar medium}
\label{sec:diffuse}

Before annihilation, \pos\ diffuse through the ISM along magnetic field lines.  \citet{jea06} have calculated the propagation distances of \pos\ in the ISM based on the cross-sections of \citet{gue05}.  They consider three regimes of propagation,  a quasi-linear regime in which \pos\ are transported in resonance with Alfv\'{e}n waves, a collisional in-flight phase in which the \pos\ are losing energy and a collisional thermalised phase.  Combining the propagation modes with the Ps formation and annihilation cross sections they calculate a maximum distance, $d_{\rm max}$, over which the \pos\ can propagate before annihilation.  It is found that in the warm phases of the ISM (8000 K) \pos\ do not propagate more than $\approx 50$ pc in a typical annihilation time of $\sim 10^{4}$ yrs.  Even those \pos\ which survive longer than the typical annihilation time do not propagate much further since the diffusion constant in the collisional regime is very small ($\sim 10^{-11}\ {\rm pc}^{2}\ {\rm yr}^{-1}$ for a warm ionised ISM).  Thus except in hot phases ($10^{6}$ K), escaping \pos\ remain local to the \pos\ source and we call this associated volume $V_{\rm source}$.

\subsubsection{Supernov\ae}
\label{sec:sne}

Supernov\ae\ (SNe) are thought to be a major, and possibly dominant, contributor to the origin of the Galactic \pos\ (\citealt{mil02}).  SNe Ia produce \pos\ from radioactive $\beta+$ decay of $^{56}{\rm Ni} \rightarrow\ ^{56}{\rm Co} \rightarrow\ ^{56}{\rm Fe}$, with a smaller contribution from $^{44}{\rm Ti} \rightarrow\ ^{44}{\rm Sc} \rightarrow\ ^{44}{\rm Ca}$, and still a lesser contribution from $^{26}{\rm Al} \rightarrow\ ^{26}{\rm Mg}$ (\citealt{cha93}).    The contribution from SNII is thought to be fewer by about an order of magnitude (\citealt{mil02}) and here we only consider SNIa.

\citet{mil99} find the average positron yield from SNe Ia is $\approx 2 \times 10^{54}$ \pos\ (see their table~4).  Of these, after 2000 days, approximately 2--6 per cent escape the supernova, whilst less than 1 per cent of the trapped \pos\ survive.  We assume that all other \pos\ annihilate.  Thus we may consider the Ps recombination line emission from SNe Ia in two components: the initial emission from the Ps formed in the supernova explosion, and the emission from the Ps formed from the escaping \pos.  In the following we assume an escape fraction of 5 per cent.

\begin{figure}
\begin{center}
\centering \includegraphics{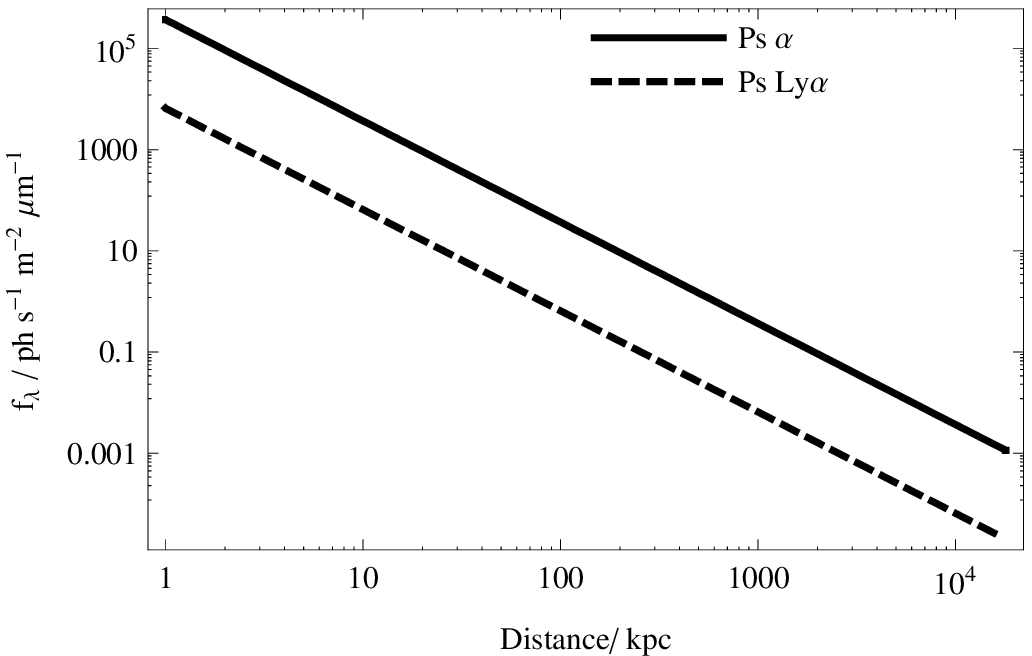}
\caption{The line flux density of \psla\ and \psa\ as a function of distance for Ps formed in the ejecta of SNe Ia in the first 2000 days, i.e.\ for those \pos\ which do not escape.  See the text for details.}
\label{fig:sneflux}
\end{center}
\end{figure}

We assume that 95 per cent of \pos\ annihilate in the first 2000 days after a type Ia SN explosion.   If we conservatively assume that this takes place evenly across the 2000 days we have a \pos\ source rate of $r\approx1.1 \times 10^{46}$ \pos\ s$^{-1}$.  We make the further simplifying assumption that all \pos\ which annihilate in the ejecta thermalise before doing so, and we assume that the temperature of the ejecta is $\sim 10^{6}$ K.  Thus using equation~\ref{eqn:flux_source} and the appropriate value from Table~\ref{tab:temp} we can calculate the line flux density.  The results are shown in Figure~\ref{fig:sneflux} for both \psa\ and \psla, as a function of distance from Galactic SNe to the distance of the Virgo cluster assuming a high visual extinction of $A_{\rm V}=5$ mag, such as would be appropriate for the Galactic centre.  The feasibility of detection is deferred until section~\ref{sec:expt}.

We now consider the fate of the \pos\ which survive and escape the ejecta.  We assume these \pos\ are thermalised in the interstellar medium, and consist of 5 per cent of original SN \pos\ yield.  \citet{mil99} show that the typical energy of an escaped \pos\ is $\approx 400$keV.  The lifetime of \pos\ in the ISM is determined by the energy of the \pos, the density of the ISM, and the cross-sections for the processes of radiative, recombination, charge-exchange and direct annihilation, etc.  These cross-sections have been calculated by \citet{gue05} for various phases of the ISM: molecular (10 K), cold (80 K), warm neutral (8000 K), warm ionised (8000 K), hot ($10^{6}$ K).


\citet{kno05} discuss the SNIa rate and distribution in the Galaxy.  Although the exact numbers are not known, estimates based on the 511keV line flux (\citealt{kno05}) and from extragalactic SNe rates (\citealt{tam94}; \citealt{cap97}; \citealt{man05}) suggest that there are 0.3--1.1 SNIa per century in the bulge, although \citet{pra06} predict a much lower value of $\approx 0.07$ SNIa per century based on extrapolations from SNIa rates in early type galaxies.  We assume a value of 0.5 SNIa per century in the bulge or a timescale of $T_{\rm SNB}=200$~yr.  Thus in a volume $V_{\rm source}$ (see \S~\ref{sec:diffuse}), there would be a typical timescale between SNe of $T_{\rm SN}=T_{\rm SNB}\frac{V_{\rm bulge}}{V_{\rm source}}$.  For typical values in the bulge of $T_{\rm ann}\approx 10^{4}$~yr and $d_{\rm max} \approx 50$~pc  at $10^{4}$ K (\citealt{jea06}), $T_{\rm SN} \gg T_{\rm ann}$ therefore we may consider each SN event individually since the probability of multiple SNe in the same volume $V_{\rm source}$ within a few $e$-folding timescales is very small.

The number of \pos\ in a volume $V_{\rm source}$ following a SN is  governed by the differential equation
\begin{equation}
\frac{{\rm d}N}{{\rm d}t}=-\frac{N(t)}{T_{\rm ann}},
\end{equation}
hence,
\begin{equation}
N(t)=N_{0}e^{-\frac{t}{T_{\rm ann}}},
\end{equation}
where $t$ is the time since the SNe and $N_{0}$ is the initial number of surviving \pos\ per SN, and the annihilation timescales, $T_{\rm ann}$ are given by \citet{jea06}.  Since we are interested in the formation of Ps we need to distinguish between the timescale for annihilation by any means, $T_{\rm ann}$, which includes Ps formation and direct annihilation, and the timescale for Ps formation alone\footnote{This is the timescale for annihilation of \emph{free} \pos\ in the ISM, and includes the timescale for the formation of Ps.  This should not be confused with the timescale for Ps annihilation,
 $T_{\rm Ps}$}.  
 The former will govern the depletion of Ps in the ISM, and the latter will govern the formation rate of Ps.
So the number of Ps formed as a function of time is therefore
\begin{equation}
N_{\rm Ps}=f_{\rm Ps} N_{0}\left(1-e^{-\frac{t}{T_{\rm ann}}}\right),
\end{equation}
where as before $f_{\rm Ps}=T_{\rm ann}/T_{\rm Ps}$ is the fraction of \pos\ which form Ps.  
Thus the Ps formation rate is
\begin{equation}
\frac{{\rm d}N_{\rm Ps}}{{\rm d}t}=\frac{f_{\rm Ps}N_{0}}{T_{\rm ann}}  e^{-\frac{t}{T_{\rm ann}}}.
\end{equation}
This expression can be substituted for $r \times f_{\rm Ps}$ in equation~\ref{eqn:flux_source} to give the total line flux for \pos\ originating from SNIa.  

\begin{figure}
\centering \includegraphics[scale=1.0]{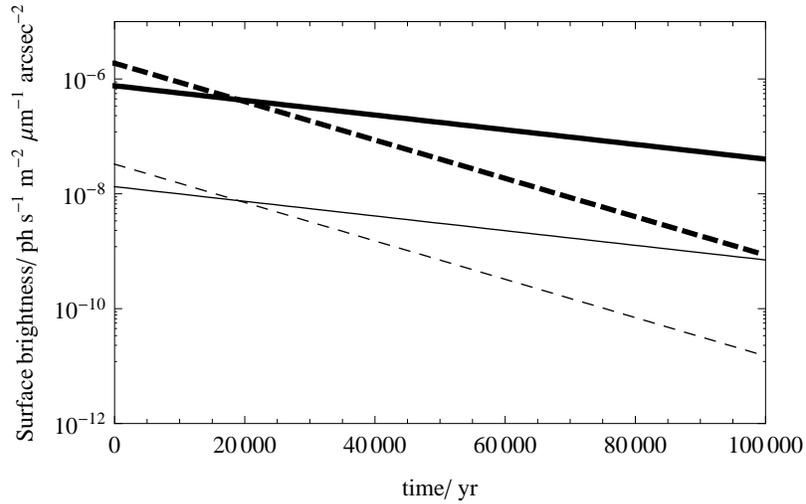}
\caption{The surface brightness of \psa\ (thick lines) and \psla\ (thin lines) emission from the \pos\ released in a SN~Ia located near the Galactic centre.  The continuous lines are for an ionised medium at 8000K and the dashed lines for a neutral medium at 8000K.}
\label{fig:sn}
\end{figure}

Using values of $T_{\rm ann}$ and $d_{\rm max}$ from \citet{jea06} (for 1Mev \pos), values of $f_{\rm Ps}$ from \citet{gue05}, $\alpha$ and $\beta$ from \citet{wal96}  (and assuming the 8000K values from \citealt{jea06} are appropriate for the $10^{4}$ K values of \citealt{wal96}), we have calculated the surface brightness of \psla\ and \psa\ from SNIa as a function of time for both a neutral and ionised ISM at $10^{4}$ K.   We assume the SN is located at the Galactic centre with a distance 8kpc and $A_{\rm V}=5$ mag.  The results are shown in Figure~\ref{fig:sn}, and are exceedingly faint --- orders of magnitude below the detection limit for any planned space or ground based mission.

These fluxes correspond to a total \pos\ annihilation rate over the entire bulge of $\approx 2 \times 10^{43}$ \pos\ s$^{-1}$, assuming one SN every 200 yrs, which is in good agreement with the values \citet{mil02} of $\approx 4 \times 10^{43}$ \pos\ s$^{-1}$ for the entire Galaxy with bulge/disc ratios of 0.2--3.3.  This annihilation rate corresponds to a total flux of 511 keV photons of $\approx 40$ photons s$^{-1}$ m$^{-2}$, which is in order of magnitude agreement with the value of 11.2 photons s$^{-1}$ m$^{-2}$ reported by \citet{kno05}.

\subsubsection{Nov\ae}
\label{sec:novae}

Nov\ae\ are potentially significant sources of positrons.  Nov\ae\ are caused by a runaway thermonuclear reaction due to the accretion of hydrogen rich material onto the surface of a white dwarf.   Positrons are expected to be produced by $\beta^{+}$ decay of radioactive nuclei produced during the runaway, and carried to the surface of the nova by rapid convection (\citealt{cla74}).  Detailed calculations of the expected \pos\ annihilation signatures were first calculated by \citet{lei87}.  These show an early peak in emission due to decay from $^{13}$N followed by decay from $^{18}$F, with lesser contributions from $^{15}$O, $^{14}$O and $^{34m}$Cl.  This early peak lasts $\lsim 2$ days,   but may be followed by longer duration, but fainter, emission from $^{22}$Na.  This is particularly relevant for ONe nov\ae\, which have a larger abundance of Na than CO nov\ae.  The longer duration 
emission is pertinent since the bright early peak occurs before the maximum in visual luminosity, i.e. before discovery (\citealt{her02}).

\begin{figure}
\centering \includegraphics[scale=1.0]{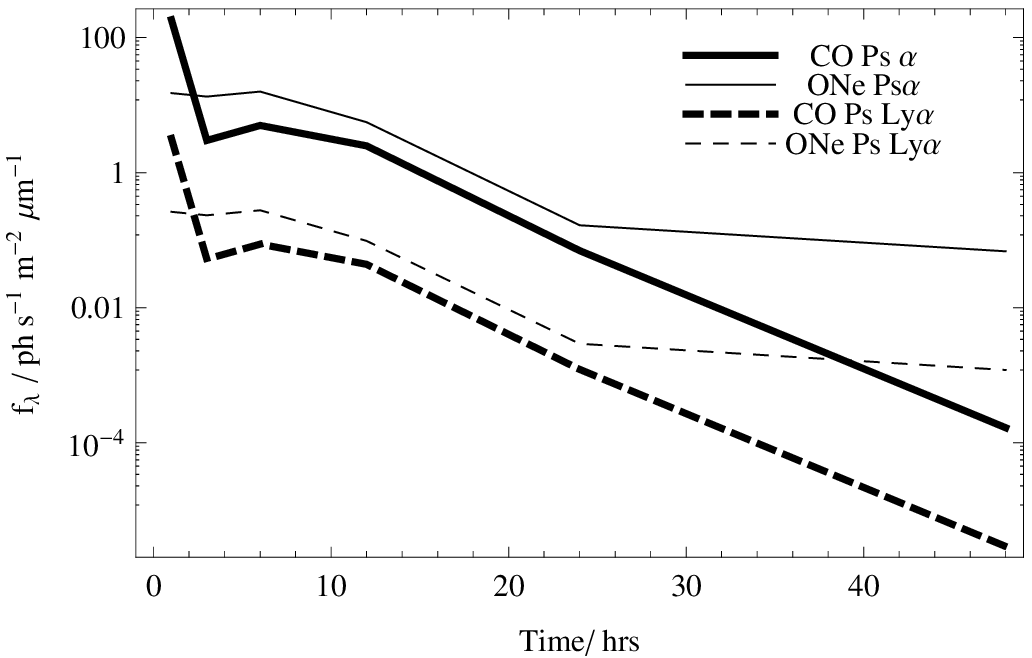}
\caption{The \psa\ and \psla\ flux densities of CO and ONe nov\ae\ at a distance of 8 kpc as a function of time.}
\label{fig:novae}
\end{figure}

\citet{her99} provide updated light curves of the expected 511 keV radiation from both CO and ONe nov\ae.  We use their table 2 of the fluxes at various epochs along with equation~\ref{eqn:wallyn} to calculate the expected \psla\ and \psa\ line strengths.  \citet{lei87} assume that the \pos\ thermalise before forming Ps, and have a temperature of $<10^{5}$K, such that Ps formation dominates over direct annihilation.  Following this prescription we assume a temperature of $10^{4}$K and an ionised medium and use the appropriate values from Table~\ref{tab:temp}.  Nov\ae\ are thought to reside mainly in the bulge of the Galaxy, therefore we assume a distance of 8kpc and a conservative extinction of $A_{\rm V}=5$ mag.  Figure~\ref{fig:novae} shows the resulting \psa\ and \psla\ fluxes for the CO and ONe nov\ae.  The flatter light curve of the ONe nov\ae\ is due to the higher abundance of $^{22}$Na.  We defer a discussion of the feasibility of detection to \S~\ref{sec:expt}.

\subsubsection{Low mass X-ray binaries and microquasars}
\label{sec:mqso}
Microquasars are a promising source of \pos\ in the Galaxy (\citealt{gue06}).  The accretion disc of a low mass X-ray binary will produce \el-\pos\ pairs, which in a microquasar are ejected into the ISM via a relativistic jet in a manner analogous to AGN (see \S~\ref{sec:agn}), though not as powerful.  \citet{gue06} have examined the expected contribution from microquasars in accounting for the Galactic Ps annihilation radiation, and have also examined misaligned microquasars, in which the jet impinges on the secondary star as a promising Ps point source.  Binary systems with colliding winds, or jet-wind interactions, can also be highly energetic sources capable of producing $>100$ MeV $\gamma$-rays (\citealt{tav09}), which are also potentially promising point sources of \pos.

There is good evidence that even those LMXBs which do not produce a jet are an important source of \pos\ in the Galaxy; there is an asymmetric distribution of 511keV radiation from the inner Galactic disc, which coincides with the asymmetric distribution of LMXBs (\citealt{wei08}).  For LMXBs without a jet, the \pos\ can escape in a stellar wind.  Since the origin of the \pos\ is the same as for microquasars, and since the annihilation timescale is long compared to the diffusion timescale, the diffuse annihilation radiation is expected to be identical for LMXBs and microquasars.  However, as discussed below microquasars have the important distinction that in some cases the jet can collide with the secondary star, producing a point source of emission.

From consideration of energetics, and from an empirical analysis of the INTEGRAL SPI data, \citet{gue06} calculate a typical \pos\ flux of $\approx 10^{41}$ \pos\ s$^{-1}$ from a microquasar.  The annihilation in the jet is expected to be small (see the discussion for AGN in \S~\ref{sec:agn}), and once the jet collides with a gas cloud, thermalisation will take place very quickly (\S~\ref{sec:agn}).  Thereafter the \pos\ will diffuse along the magnetic field of the ISM as described in \S~\ref{sec:diffuse}, and will eventually annihilate as described in \S~\ref{sec:spec}.  Thus the Ps recombination line emission from ejected \pos\ from microquasars and LMXBs will form an extended emission region $\sim50$pc across.  Assuming the Ps recombination line emission is uniform over the distance $d_{\rm max}$ given by \citet{jea06} we have calculated the surface brightness of a cloud using the same assumptions for $f_{\rm Ps}$, $\alpha$, $\beta$ and  as in \S~\ref{sec:sne}.  The resulting surface brightness of the escaped \pos\ from microquasars is very faint, and are listed in Table~\ref{tab:mqso}.

\begin{table}
\center
\caption{The surface brightness of \psa\ and \psla\ emission resulting from \pos\ escaping into the ISM from the jets of microquasars or winds of LMXBs, see the text for details.}
\label{tab:mqso}
\begin{tabular}{lll}
& \multicolumn{2}{c}{Surface brightness/ ph s$^{-1}$ m$^{-2}$ arcsec$^{-2}$ $\mu$m$^{-1}$} \\
T/ K & \psla & \psa \\ \hline
10& $3 \times 10^{-7}$ &$3 \times 10^{-5}$ \\
80& $1 \times 10^{-7}$ &$8 \times 10^{-6}$ \\
8000 (neutral)& $1 \times 10^{-8}$ &$8 \times 10^{-7}$ \\
8000 (ionised)& $1 \times 10^{-8}$ &$8 \times 10^{-7}$ \\
1000000& $2 \times 10^{-14}$ &$1 \times 10^{-12}$ \\
\end{tabular}
\end{table}

\citet{gue06} also consider misaligned microquasars, in which the jet from the accretion disc impinges on the secondary star, as a possible class of point sources of  annihilation radiation.  Their model accounts for deformation of the stellar surface due to the jet, and the fraction of annihilation photons which escape the atmosphere of the star and the jet induced cavity in the direction of the observer.  Thus they calculate expected 511keV fluxes for several confirmed and candidate misaligned microquasars.  

We have converted these fluxes into the equivalent recombination line fluxes using the assumptions implicit in equation~\ref{eqn:wallyn} with the further assumption that
recombination line photons emitted in a upward direction escape the surface of the star.  This assumption is reasonable since the interaction depth at which relativistic \pos\ from the jet are slowed down enough to form Ps via charge exchange is $\lambda_{\rm form} \sim 0.1$ g cm$^{-2}$ (\citealt{gue06}), whereas the mean free path of visible photons in a typical stellar photosphere is $\lambda_{\rm 5000} \sim 3.8$ g cm$^{-2}$, and assuming this value is valid for the wavelengths of both \psla\ and \psa\ upward photons should escape the star. 
%
%
%
The resulting fluxes are given in Table~\ref{tab:misaligned_mqso} for a $10^{4}$ K and a $10^{6}$ K gas.  We assume the visual extinction corrections, $A_{\rm V}$, from \citet{schl98}, but note these may not be accurate for sources close to the Galactic plane.

\begin{table}
\center
\caption{The \psla\ and \psa\ fluxes for the microquasars of \protect\citet{gue06} as described in \S~\ref{sec:mqso}.  The top two objects are confirmed mis-aligned microquasars, the remaining objects are potentially mis-aligned.    Fluxes are given in units of ph s$^{-1}$ m$^{-2}$ $\mu$m$^{-1}$.}
\label{tab:misaligned_mqso}
\begin{tabular}{lllll}
& \multicolumn{2}{c}{$10^{4}$ K} & \multicolumn{2}{c}{$10^{6}$ K} \\
Source & \psla & \psa & \psla & \psa \\ \hline
GRO J1655-40 & 249 & 73.8 & 9.25 & 2.70 \\
 V4641 Sgr & 62.7 & 8.74 & 2.33 & 0.320 \\
 XTE J1550-564 & 51.3 & 24.7 & 1.90 & 0.903 \\
 XTE J1118+480 & 3810 & 299 & 141 & 11.0 \\
 LSI+61$^{\circ}$ 303 & 302 & 166 & 11.2 & 6.06 \\
 Cyg X-1 & 134 & 99.4 & 4.99 & 3.64 \\
 Sco X-1 & 997 & 114 & 36.9 & 4.19 \\
 LS5039 & 2.27 & 39.8 & 0.0843 & 1.46 \\
 SS433 & 11.5 & 22.8 & 0.425 & 0.835 \\
 GRS 1758-258 & 0.114 & 4.10& 0.00423 & 0.150 \\
 Cyg X-3 & 0.386 & 4.41 & 0.0143 & 0.161 \\
 CirX-1$_{2}$ & 0.0325 & 2.52 & 0.00121 & 0.0923 \\
 GRS 1915+105$_{2}$ & 0.0000398 & 0.546 & 1.48$\times 10^{-6}$ & 0.0200 \\
 GX 339-4 & 48.2 & 12.4 & 1.79 & 0.454 \\
 XTE J1748-288 & 1.33$\times 10^{-21}$ & 0.00243 & 4.94$\times10^{-23}$ & 0.0000891\\
 1E 1740.7-2942 & 2.62$\times 10^{-27}$ & 0.000251 & 9.72$\times 10^{-29}$ &  9.20$\times 10^{-6}$ \\
 IGR J17091-3624 & 109 & 12.7& 4.05 & 0.464 \\
 IGR J17303-0601 & 71.2 & 11.8 & 2.64 & 0.433 \\
 IGR J17464-3213 & 0.934& 5.80 & 0.0346 & 0.212 \\
 IGR J18406-0539 & 9.83$\times 10^{-10}$ & 0.194 & 3.65$\times 10^{-11}$ & 0.00711
\end{tabular}
\end{table}

\subsubsection{Pulsars}
\label{sec:pulsars}

Pulsars produce \pos\ via pair-production due to the interaction of an \el\ with the strong magnetic field.  We will mainly consider the Ps formation rates for \pos\ escaping the pulsar, but first we briefly consider \pos\ forming Ps in the atmosphere of the pulsar.  

\citet{uso96} describe a model in which Ps is formed from the decay of synchroton $\gamma$-ray radiation produced in the strong magnetic fields of pulsars.  It is argued that the Ps escape from the polar gap with Lorentz factors of $\Gamma \sim 10^{7}$, and thereafter quickly dissociate.  The Lorentz factors for the Ps in the pulsar are thus so high that the emission lines will be very highly redshifted for any orientation of the beam, even toward or tangential to the observer, due to time dilation.  \citet{bar01} argue that Ps will be short lived in the atmosphere of pulsars except in the case of very high Lorentz factors or for a narrow range of magnetic field strengths.  Thus we will not consider direct observation of Ps in the atmospheres of pulsars further.

We now turn our attention to Ps formed from \pos\ escaping the pulsar.  \citet{stu79} estimate the \pos\ production rate from the Crab pulsar to be $\approx 2 \times 10^{41}$ \pos\ s$^{-1}$.  \citet{wan06} examine pair-production from three separate catogories of pulsars: normal pulsars (e.g.\ the Crab pulsar), magnetars in gamma-ray burst (GRB) progenitors, and milli-second pulsars.  They estimate \pos-\el\ injection rates of $10^{41}$ \pos\ s$^{-1}$ and $10^{39}$ \pos\ s$^{-1}$ for the Crab and Vela pulsars respectively.  For millisecond pulsars they estimate an injection rate of $5 \times 10^{37}$ \pos\ s$^{-1}$.  We do not consider the magnetar GRB progenitors here, since although they may contribute to the diffuse \pos\ annihilation radiation they are too infrequent in the Galaxy ($\approx 3\times10^{-6}$ yr$^{-1}$; \citealt{pir04}) to be a tenable observational target.  

We assume that the \pos\ leaving the pulsar thermalise before forming Ps, thus they will first diffuse through the ISM according to \S~\ref{sec:diffuse}.  The number of positrons as a function of time in the volume around the source, $V_{\rm source}$ (see \S~\ref{sec:diffuse}), will be
\begin{equation}
N(t)=n_{\rm e^{+}} t e^{-\frac{t}{T_{\rm ann}}},
\end{equation}
where $n_{\rm e^{+}}$ is the positron injection rate and $T_{\rm ann}$ is the annihilation timescale, and we have made the simplifying assumption that \pos\ thermalise instantly on injection into the ISM.  Thus the number of Ps atoms formed as a function of time is
\begin{equation}
N_{\rm Ps}=f_{\rm Ps} n_{\rm e^{+}} t (1 - e^{-\frac{t}{T_{\rm ann}}})
\end{equation}
where as before $f_{\rm Ps}$ is the fraction of \pos\ which form Ps before annihilation.  Therefore the Ps formation rate in the volume $V_{\rm source}$ around a pulsar is
\begin{equation}
\frac{{\rm d}N_{\rm Ps}}{{\rm d}t}=n_{\rm e^{+}} f_{\rm Ps} \left(1-e^{-\frac{t}{T_{\rm ann}}} +\frac{e^{-\frac{t}{T_{\rm ann}}} t }{T}\right),
\end{equation}
which in the limit $t>>T_{\rm ann}$ becomes
\begin{equation}
\label{eqn:pulsar_ps}
\frac{{\rm d}N_{\rm Ps}}{{\rm d}t}\approx n_{\rm e^{+}} f_{\rm Ps}.
\end{equation}

Thus assuming a positron injection rate of $10^{41}$ \pos\ s$^{-1}$ for the Crab pulsar and following the same assumptions as in \S~\ref{sec:sne} we can calculate the Ps surface brightnesses for \psa\ and \psla\ around pulsars.  The results are given in Table~\ref{tab:ps_pulsars}; the surface brightnesses for escaped \pos\ from pulsars are extremely faint.

\begin{center}
\begin{table}
\caption{The surface brightness of the diffuse \psa\ and \psla\ emission arising from \pos\ escaping from pulsars as a function of temperature.  For details see \S~\ref{sec:pulsars}.}
\label{tab:ps_pulsars}
\begin{center}
\begin{tabular}{lll}
Temp/ K & \multicolumn{2}{c}{Surface brightness/} \\
& \multicolumn{2}{c}{ph s$^{-1}$ m$^{-2}$ $\mu$m$^{-1}$ arcsec$^{-2}$}\\
& \psla\ & \psa\ \\ \hline
10 & $6 \times 10^{-8}$ & $3 \times 10^{-5}$\\
80 & $2 \times 10^{-8}$ & $8 \times 10^{-6}$ \\
8000 (neutral) & $3 \times 10^{-9}$ &$8 \times 10^{-7}$\\
8000 (ionised)  & $3 \times 10^{-9}$  & $ 8\times10^{-7}$ \\
1000000  & $3 \times 10^{-15}$ & $1\times10^{-12}$
\end{tabular}
\end{center}
\end{table}
\end{center}


\subsection{Extragalactic sources}

The only detected sources of Ps annihilation radiation are from our own Galaxy and the Sun.  However the production processes responsible for the generation of \pos\ within the Galaxy will also be active in other galaxies, and hence in principle they too should be sources of Ps recombination line emission.  This is discussed in \S~\ref{sec:galaxies}.  The pair production processes in jets discussed in \S~\ref{sec:mqso} will be applicable to the jets of AGN, in which the much larger energies will produce many more pairs, as discussed in \S~\ref{sec:agn}.
In addition to these sources it is possible that very bright individual sources, such as SNe and GRBs, which are too rare to be observed in our own Galaxy, may be detected in other galaxies.

Observing extragalactic sources of Ps presents its own problems, most apparently the distance to the sources and the corresponding decrease in flux.  However, there are some advantages.  The combined Ps output of all sources in a galactic bulge will be observable together, providing an increase in the intrinsic luminosity of the Ps compared to individual Galactic sources.  Sources can be selected such that the \psa\ emission, which lies in the infrared region is not coincident with bright atmospheric emission lines.

\subsubsection{Active galactic nuclei jets}
\label{sec:agn}

Powerful relativistic jets are a property of many AGN.   These jets emit strongly at radio wavelengths with a power-law spectrum which is interpreted as being due to synchrotron emission from \el\ carried at relativistic energies along the magnetic field lines emanating from the poles of the AGN.  These jets must be electrically neutral otherwise the potential difference induced by the jet would impede and eventually halt the jet.

It is an unsolved issue whether the positive component of jets consists of protons, positrons or a mixture of both.  From a theoretical viewpoint it is expected that AGN jets should contain some \el-\pos\ pairs; a pair plasma has the advantage of explaining $\gamma$-ray jets (\citealt{bla95}) and the very high Lorentz factors ($\Gamma \gsim 5$) required to account for superluminal bulk velocities of jets (\citealt{beg84}).
From an observational view point the situation is unclear as most attempts to answer this question have had to rely on indirect methods of searching for the presence of \pos.  For example estimates on the bulk kinetic energy contained in jets on a variety of scales have been used to argue for both \el-p plasma (\citealt{cel93}) and \el-\pos\ plasma  (e.g.\ \citealt{rey96}; \citealt{war98}; \citealt{hir05}). 

\citet{mar07} searched for the 511 keV annihilation signature in 3C 120, and their non-detection places upper limits on the \pos\ content of jets (albeit not significantly constraining).  In doing so they have also developed theoretical arguments for the \pos\ content of jets which we will use here to estimate the Ps formation rate due to jets.   \citet{mar07} show that the flux of \pos\ in a jet is given by
\begin{equation}
F(e^{+})=\pi R_{0}^{2} c f \Gamma \frac{N_{0}}{2 s} E_{\rm min}^{-2 s}
\end{equation}
where $R_{0}$ is the cross-sectional radius of the jet, $f$ is the fraction of the jet which consists of a pair plasma (i.e.\ $0.5\times f$ is the \pos\ fraction of the jet), $\Gamma$ is the Lorentz factor for the bulk velocity of the jet, $E_{\rm min}$ is the minimum energy of the power-law distribution of relativistic \el-\pos, $s$ is the spectral index of the synchrotron radiation from the jet ($F_{\nu} \propto \nu^{-s}$) and $N_{0}$ is the normalisation of the 
 \el-\pos\ number density per unit energy  such that
\begin{equation}
N(E)=N_{0} E^{-2s - 1}.
\end{equation}

The parameters $N_{0}$, $R_{0}$, $E_{\rm min}$, $s$ and possibly $f$ should be determined independently for individual AGN (e.g.\ as in \citealt{mar07} for 3C 120).  \citet{mar83} gives formul\ae\ to relate $N_{0}$ to observational parameters, viz.\ the angular size of the source,  $\theta$;  the frequency of synchrotron self-absorption turnover, $\nu_{\rm m}$;  the flux density at $\nu_{\rm m}$, $F_{\nu}(\nu_{\rm m})$; the spectral index $s$; the redshift $z$; the luminosity distance $D_{\rm L}$; and the Doppler beaming factor $\delta=\Gamma(1-\beta \cos \phi)^{-1}$ ($\phi$ is the angle of the velocity vector away from the line of sight and $\beta$ is the velocity divided by $c$).

\citet{ghi93} provide measurements of the above observable parameters for different classes of AGN, viz.\ BL Lacs, core-dominated quasars (low and high polarisation), lobe dominated quasars and radio galaxies.  We have followed the formul\ae\ of \citet{mar83} to calculate $N_{0}$ assuming a spectral index $s=0.75$ and median values of  $\Gamma$,  since these measurements are not available for all the sources.  With the further assumption that $E_{\rm min}=m c^{2}$ (i.e.\ $\gamma_{\rm min}=1$) and $f=0.5$, and setting $R_{0}=7.4 \times 10^{13}$m (as calculated for 3C 120 by \citealt{mar07}), we can calculate the positron flux for all the AGN in \citet{ghi93}.  We report the mean, median and standard deviation of the fluxes calculated for each class of source in Table~\ref{tab:agnposflux}.  For the sources of \citet{ghi93} the blazars are the most powerful \pos\ emitters.  Blazars also have the advantage that the jet is aligned along the line of sight of observations, and is thus not obscured by the dusty torus which surrounds the AGN.  Therefore we focus our attention on blazars as the most promising AGN candidate for a successful Ps detection and we assume the mean value of $10^{49}$ \pos\ s$^{-1}$ as the default flux.  We note that the assumption of a uniform jet implicit in these calculations may result in the \pos\ flux being too small by a factor $\sim 10$  (cf.\ \citealt{mar07}; see also a similar discussion for uniform/ non-uniform wind \citealt{mar77}).   Furthermore there are significant uncertainties in the parameters taken from \citet{ghi93}, particularly the self-absorption frequency and consequently the flux at the self-absorption frequency for which it was assumed that the frequency of the VLBI observation is equal to the self-absorption frequency.  Bearing these caveats in mind, the values in Table~\ref{tab:agnposflux} should be treated as approximate.

\begin{center}
\begin{table}
\caption{Average positron fluxes calculated from the AGN observations of \protect\citealt{ghi93}.}
\label{tab:agnposflux}
\begin{tabular}{lllll}
Class & No.\ of sources & Mean/ \pos/ s& Median/ \pos/s & $\sigma$/ \pos/s \\ \hline
BL Lac objects & 22 &  $10^{49} $ & $10^{47}$ & $10^{49}$ \\
Core-dominated high-polarisation quasars & 24 & $10^{46} $ &  $10^{46} $ & $10^{46} $\\
Core-dominated low-polarisation quasars & 24 & $10^{47} $ &  $10^{46} $ & $10^{48} $\\
Lobe-dominated quasars & 11 & $10^{48}$ & $10^{47}$ & $10^{48}$ \\ 
Radio galaxies & 8 & $10^{47}$ & $10^{46}$ & $10^{47}$ \\
\end{tabular}
\end{table}
\end{center}

Having calculated the \pos\ flux from AGN jets we are now in a position to estimate the Ps flux.  We assume that the Ps formation rate in the jet is insignificant since the \pos\ have energies above the ionisation energy of Ps.  Furthermore \citet{fur02} show that annihilation before thermalisation occurs for only $\sim 5$ per cent of \pos\ produced in jets.  When the jet interacts with a gas cloud, the positrons will thermalise on a timescale,
\begin{equation}
T_{\rm therm}=4.8 \times 10^{3} \left(\frac{k_{\rm B}T}{1\ {\rm keV}}\right)^{\frac{3}{2}} \left(\frac{10^{-3}\ {\rm cm}^{-3}}{n_{\rm e}}\right) {\rm yr}
\end{equation}
(\citealt{fur02}), which for $T=10^{4}$ K and $n_{\rm e}=10^{3}$ cm$^{-3}$ is 0.12 yr.  Thus we assume that \pos\ thermalise instantly on interaction with a cloud.

\begin{figure}
\begin{center}
\subfigure[\psla]{
\centering \includegraphics{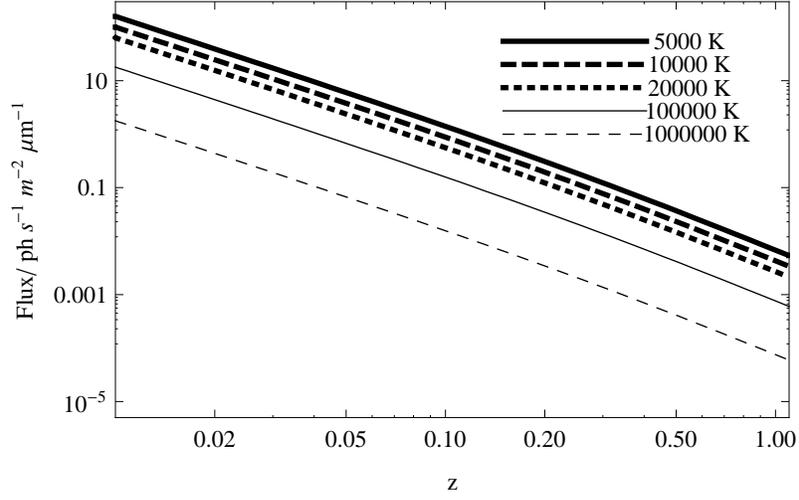}
}\\
\subfigure[\psa]{
\centering \includegraphics{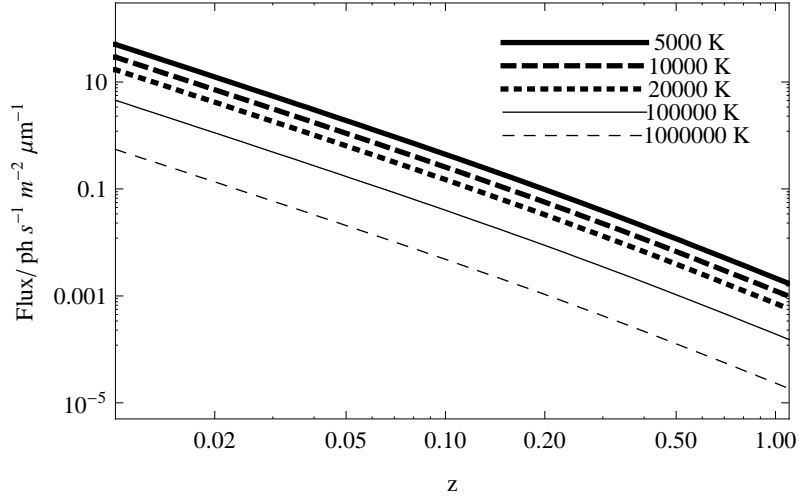}
}
\caption{The line flux density of \psla\ and \psa\ for typical AGN jets as a function of redshift for different temperatures in an ionised ISM.  The assumptions in calculating the fluxes are described in the text.}
\label{fig:psjets}
\end{center}
\end{figure}

After thermalisation the \pos\ diffusion distance will be very small (cf.\ \S~\ref{sec:diffuse}; \citealt{jea06}), thus the interaction of a jet with a cloud will produce a bright spot of Ps formation.  Thus assuming $f=10^{49}$ \pos\ s$^{-1}$ and $A_{\rm V}=1$ mag we have calculated the Ps flux for \psla\ and \psa\ as a function of redshift, and the results are plotted in Figure~\ref{fig:psjets}.

%
%
%
%
%
%
%

\subsubsection{Galaxies}
\label{sec:galaxies}

\begin{figure}
\centering \includegraphics[scale=0.8]{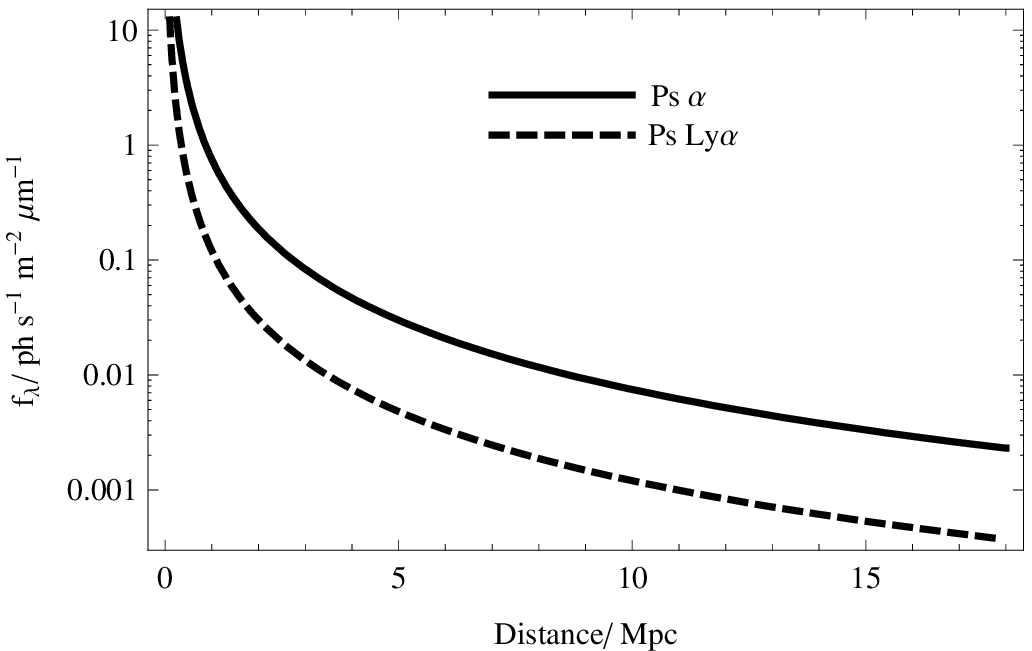}
\caption{The line flux density of \psa\ and \psla\ in spiral galaxies as a function of distance.}
\label{fig:galaxy}
\end{figure}

The total 511keV emission from the Galactic bulge and disc is $10.5 \pm 0.6$ and $7 \pm 4$ ph s$^{-1}$ m$^{-2}$ respectively (\citealt{kno05}).  Using equation~\ref{eqn:wallyn} and assuming a temperature of $10^{4}$ K and a visual extinction of $A_{\rm V}=2$ mag this translates into a line flux density of $f_{\lambda}\approx12000$ and 1900 ph s$^{-1}$ m$^{-2}$ $\mu$m$^{-1}$ for \psa\ and \psla\ respectively.  We assume that this value holds for all other large spiral galaxies.  Thus for local galaxies we can compute the expected line flux density by scaling the above numbers by $(8/d)^{2}$, where $d$ is the distance to the galaxy in kpc; we show results out to a distance of the Virgo cluster ($\sim$18 Mpc) in Figure~\ref{fig:galaxy}.

We note that the assumption of the similarity of other galaxies to the Milky Way in terms of \pos\ production and annihilation can only be considered an approximate estimate; different classes of galaxies could produce different amounts of \pos\ than the Milky Way.  In particular starbursts and ULIRGs could be copious emitters due to the SNe and other high energy processes associated with star-formation regions.

\subsubsection{Supernov\ae\ and gamma-ray bursts}

Supernov\ae\ and gamma ray bursts are both expected to produce copious amounts of positrons (see \S~\ref{sec:sne}).  They are too rare to be useful candidates of \emph{Galactic} Ps, and their remnants are too diffuse, but they could be very promising candidates  for observation in other galaxies..  Figure~\ref{fig:sneflux}  shows the estimated line flux density for SNIa as a function of distance.  

The results are expected to be similar for GRBs.  \citet{fur02b} estimate $\approx 6 \times 10^{54}$ \pos\ to be released in a GRB, with only a ``small fraction'' annihilating in the early stages during which the ejecta expands (cf. 5 per cent for SNeIa in \S~\ref{sec:sne}).  Similarly \citet{cas04} predict $\sim 10^{54}$ \pos\ to be produced by a GRB.

\section{PROSPECTS FOR DETECTION}
\label{sec:expt}

Having examined potential sources of Ps recombination lines, and estimated the strength of the sources we now assess the feasibility of observation.  We begin with a summary of the expected Ps fluxes with a comparison to current upper limits based on the non-detections of Ps Paschen-$\beta$ by \citet{pux96}, the non-detection of Galactic 511keV point sources by \citet{kno05} and the non-detection of 511keV annihilation in the jets of 3C 120 by \citet{mar07}.  We then examine the sensitivity of current and upcoming instruments to  \psla\ and \psa\ emission.  We finish by simulating observations of \psa\ in AGN.

\subsection{Current observational limits to Ps recombination line emission}
\label{sec:current_limits}

\subsubsection{\citet{pux96}}
The most direct observational constraint on Ps recombination emission line strengths comes from the non-detection by \citet{pux96} who searched for Ps Paschen-$\beta$ from the Galactic centre.  They found an upper-limit to the line strength of $3 \times 10^{-19}$ W m$^{-2}$.  If we assume a temperature of the ISM of $10^{4}$ K and an extinction of $A_{\rm V}=5$ mag, we can use the relative line strengths predicted by \citet{wal96} to convert this to a line strength of \psla=$19$ and \psa=$6200$ ph s$^{-1}$ m$^{-2}$ $\mu$m$^{-1}$, where we have also divided by the line width as given by equation~\ref{eqn:linewidth}.  These upper-limits are not constraining compared to the expected fluxes calculated in \S~\ref{sec:sources}.

\subsubsection{\citet{kno05}}
We have discussed the limits on the Ps recombination emission line strength provided by current 511keV observations as given by \citet{kno05} in \S~\ref{sec:galactic}.  For instance, if we assume that there are 100 sources of equal strength responsible for the total Galactic 511keV radiation then this equates to line strengths of \psla=$0.02$ and \psa=$6$ ph s$^{-1}$ m$^{-2}$ $\mu$m$^{-1}$, see Figure~\ref{fig:pointsources}.

\citet{kno05} also provide 511keV  3$\sigma$ flux limits for individual sources (see their table 4).  We convert these to \psla\ and \psa\ flux limits again assuming $T=10^{4}$ K and using the appropriate extinction values from \citet{schl98} or $A_{\rm V}=5$ mag for sources close  the Galactic plane for which the extinction models are inaccurate.  We use the most constraining observations for each class of source and compare these to our predictions in Table~\ref{tab:source_summary}.  From our predictions it is not expected that any of the sources would have been observed.

\subsubsection{\citet{mar07}}

\citet{mar07} searched for annihilation radiation in the jet of 3C 120 and find a 2$\sigma$ upper limit of $0.33$ ph s$^{-1}$ m$^{-2}$.  We again convert to Ps emission line strengths assuming $T=10^{4}$ K with $A_{\rm V}=1$ mag, and find   \psla=$0.89$ and \psa=$1.5$ ph s$^{-1}$ m$^{-2}$ $\mu$m$^{-1}$.

\subsubsection{Comparison of observational limits to expected fluxes}

We summarise the results of \S~\ref{sec:sources} in Table~\ref{tab:source_summary} and compare the results to the upper limits discussed above.  For the sources for which we gave surface brightnesses (e.g.\ supernov\ae) we convert to the flux over the entire object for comparison to the upper limits.  We also convert the fluxes predicted in \S~\ref{sec:sources} to the appropriate distance and extinction for the source in question.  We quote the results as flux densities (per $\mu$m$^{-1}$), which takes into account the width of the line and makes a comparison to the background more meaningful.  From our estimates of the fluxes of the sources it is not expected that any of the sources should have been detected in the 511keV observations.

\begin{table}
\begin{center}
\caption{A comparison of the predicted fluxes for different classes of sources in \S~\ref{sec:sources} with the upper limits discussed in \S~\ref{sec:current_limits}.  The reference for each upper limit is (a) \citet{kno05}, (b) \citet{mar07}.}
\label{tab:source_summary}
\begin{tabular}{llllll}
Source & \multicolumn{2}{c}{Predicted Flux} & \multicolumn{2}{c}{Upper limit}\\
& \multicolumn{4}{c}{ph s$^{-1}$ m$^{-2}$ $\mu$m$^{-1}$} \\
& \psla\ & \psa\ & \psla\ & \psa\ \\ \hline
Kepler SNR & 1.6 & 7.9 & 12$^{a}$ & 60$^{a}$ \\
LMXB GX 5-1 & 0.065 & 3.7 & 0.65$^{a}$ & 37$^{a}$ \\
Crab Pulsar & 130 & 110 & 190$^{a}$ & 160$^{a}$ \\ 
3C 120 & 8.9 & 1.5 & 170$^{b}$ & 50$^{b}$ \\
\end{tabular}
\end{center}
\end{table}

\subsection{Expected sensitivity of current and future instruments}
\label{sec:expected}

We now examine the sensitivity of current and future instruments to Ps emission line radiation.  The signal to noise ratio is given by
\begin{equation}
\label{eqn:snr}
SNR=\frac{f_{\lambda} t \epsilon}{\sqrt{f_{\lambda} t \epsilon+ B t \epsilon + D t + R^{2}}}
\end{equation}  
where $t$ is the exposure time, $\epsilon$ is a factor which takes into account the efficiency of the instrument and collecting area of the telescope, $B$ is the background, $D$ is the detector dark current and $R$ is detector read out noise.
Since the expected Ps recombination line emission is faint in all cases it is likely that the background will also include continuum emission from the source in question as well as the usual night sky background etc.

We first address the issue of whether it is more efficient to observe \psla\ or \psa.  \psla\ is always intrinsically brighter than \psa.  It does not follow however, that it is more efficient to observe \psla\ than \psa.  This is because the interstellar extinction, the line width, the background and the seeing all very with wavelength.

\begin{figure}
\centering \includegraphics[scale=1.0]{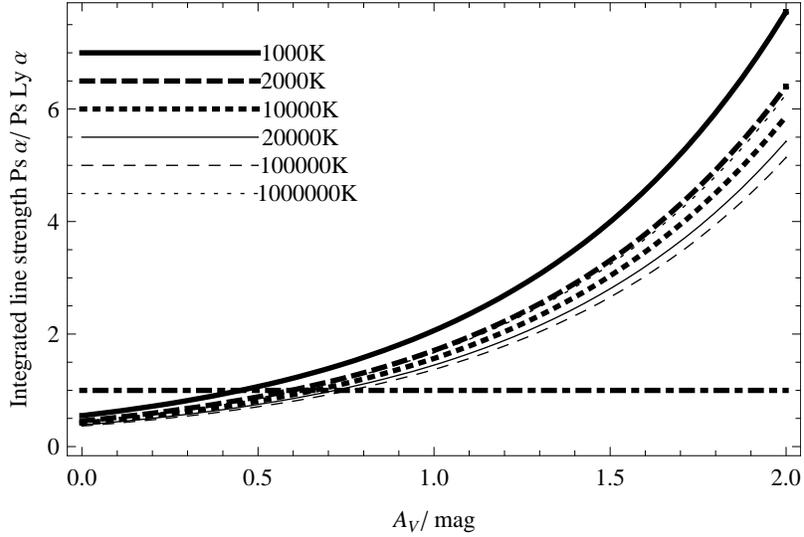}
\caption{The ratio of \psa/\psla\ for integrated line strengths as a function of visual extinction for different temperatures in the ISM.  The horizontal dot-dashed line marks the point at which the integrate line strength of both lines are equal.  Above the dashed line \psa\ is brighter than \psla.}
\label{fig:extinct_total}
\end{figure}

\begin{figure}
\centering \includegraphics[scale=1.0]{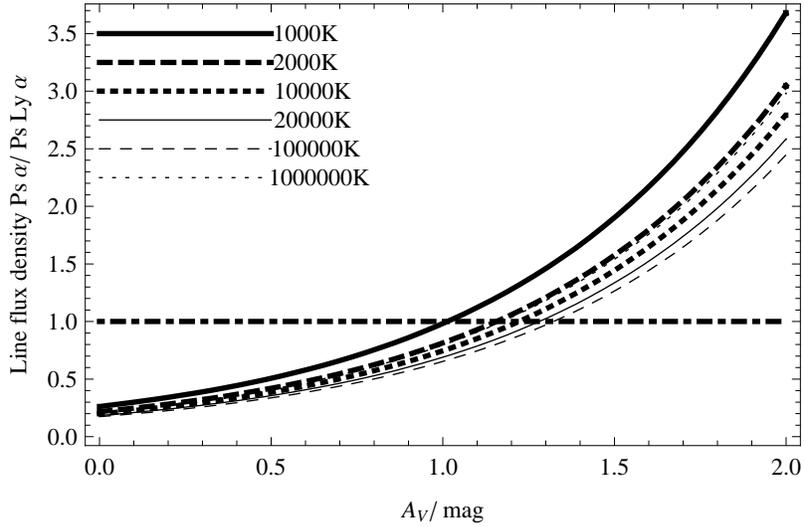}
\caption{The ratio of \psa/\psla\ for line flux densities as a function of visual extinction for different temperatures in the ISM.  The horizontal dot-dashed line marks the point at which the integrate line strength of both lines are equal.  Above the dashed line \psa\ is brighter than \psla.}
\label{fig:extinct_density}
\end{figure}

Interstellar extinction is greater for \psla\ than for \psa.  This means that the integrated line strength for \psa\ is greater than that for \psla\ for visual extinctions $A_{\rm V} \gsim 0.65$ mag, where the exact value depends on the temperature of the ISM in which the Ps is forming, see Figure~\ref{fig:extinct_total}.  However, the line width is also greater for \psa\ compared to \psla\ so the line flux density, $f_{\lambda}$, which is more meaningful to compare to the background is greater for \psa\ for visual extinctions $A_{\rm V} \gsim 1.2$ mag, see Figure~\ref{fig:extinct_density}.

A proper assessment of the most fruitful observing strategy depends on other factors, most notably the background emission and the efficiency of the instruments.  Therefore we now discuss the different backgrounds for \psla\ and \psa, and then determine the expected sensitivity of observations compared to the expected brightnesses of example objects.

The U band background at a typical good observing site is $U=21.5$ mag arcsec$^{-2}$, which is equivalent to a background of $B_{\rm U}=140$ ph s$^{-1}$ m$^{-2}$ $\mu$m$^{-1}$ arcsec$^{-2}$, which we assume is also appropriate for observations at 2430\AA.

\begin{figure}
\centering \includegraphics[scale=0.5]{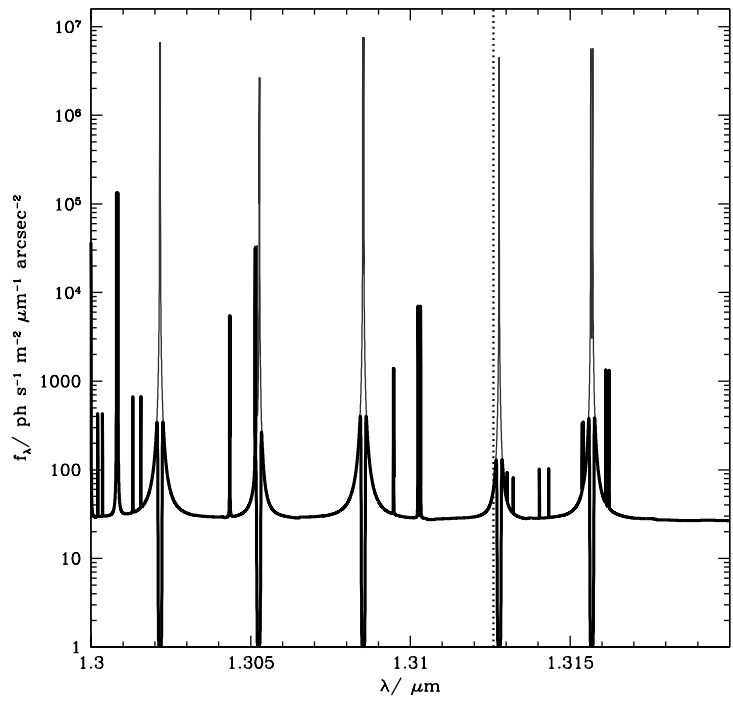}
\caption{The background around the \psa\ line is shown by the thin grey line.  The flux is logged to emphasise the fainter lines and the interline continuum.  The dashed line marks the position of the \psa\ line.  The thick black line shows the background spectrum after OH suppression.}
\label{fig:nirback}
\end{figure}

The background in the near-infrared is dominated by atmospheric hydroxyl emission lines.  Thus the background depends sensitively on the wavelength and the resolution of the observations.  Figure~\ref{fig:nirback} shows the logged background around the \psa\ line, taken from the model of \citet{ell08}.  There is a very bright OH emission line very close to the \psa\ line.  Note that Figure~\ref{fig:nirback} is at the intrinsic resolution of the night sky, i.e.\ the width of the lines is the natural width due Doppler and pressure broadening; when observed with a spectrograph, the resolution and scattering of the spectrograph will cause the OH line to severely contaminate the \psa\ line.   Subtracting the OH line results in a high Poisson and systematic noise (see \citealt{dav07}; \citealt{ell08}). Indeed, we postulate that this is the reason that \psa\ has never been observed serendipitously.

The brightness of the near-infrared sky is problematic for many areas of observational astronomy, as exemplified by the particular science case of detecting \psa.  For this reason there have been significant technological developments aimed at tackling this problem.  The implementation of several of these projects is now imminent, promising much darker near-infrared skies in the near future.  These technological developments are the motivation behind our renewed interest in the long standing observational challenge of detecting Ps recombination emission lines.

There are two technologies in particular which should be very beneficial for the detection of \psa.  These are the James Webb Space Telescope\footnote{JWST http://www.jwst.nasa.gov/} (see e.g.\ \citealt{gar06}) and OH suppressing fibre Bragg gratings (\citealt{bland04,bland08}).  The James Webb Telescope mission is designed specifically to achieve very low infrared backgrounds, both for imaging and moderate resolution spectroscopy  ($R\approx 1000$--3000).  

 \citet{bland04} and \citet{bland08} present a novel photonic solution to the problem of the NIR night sky for ground based telesopes.  Used in conjunction with adaptive optics and large telescopes OH suppression offers very low backgrounds; fibre Bragg gratings can suppress the OH lines by factors of $\approx 30$dB at a resolution 10,000.  For an estimate of the performance and impact of FBG OH suppression on astrophsyical observations see \citet{ell08}; for a demonstration of their performance in on-sky tests see \citet{bland09}.  
 
Thus future NIR spectrographs will benefit from much lower backgrounds.  The thick black line in Figure~\ref{fig:nirback} shows the expected background around the \psa\ line after suppression with FBG technology based on current performance.
We note that at NIR wavelengths the background can be further reduced with adaptive optics which permits the use of a smaller aperture to collect the same object flux.
After OH suppression the background flux density  at the wavelength of \psa\ is less than that of \psla, assuming the interline continuum in the model of \citet{ell08} is correct.  

The calculation of the difference in backgrounds is complicated by the fact that the line width of \psa\ is greater than that of \psla\ and so the background must be integrated over a larger wavelength range for \psa.  Conversely the seeing is smaller at longer wavelengths, and so the \psa\ background can be measured from a smaller aperture than for \psla; this effect is even more pronounced with adaptive optics which can further reduce the necessary aperture in the infrared, but not in the ultraviolet.

\begin{figure}
\centering \includegraphics[scale=1]{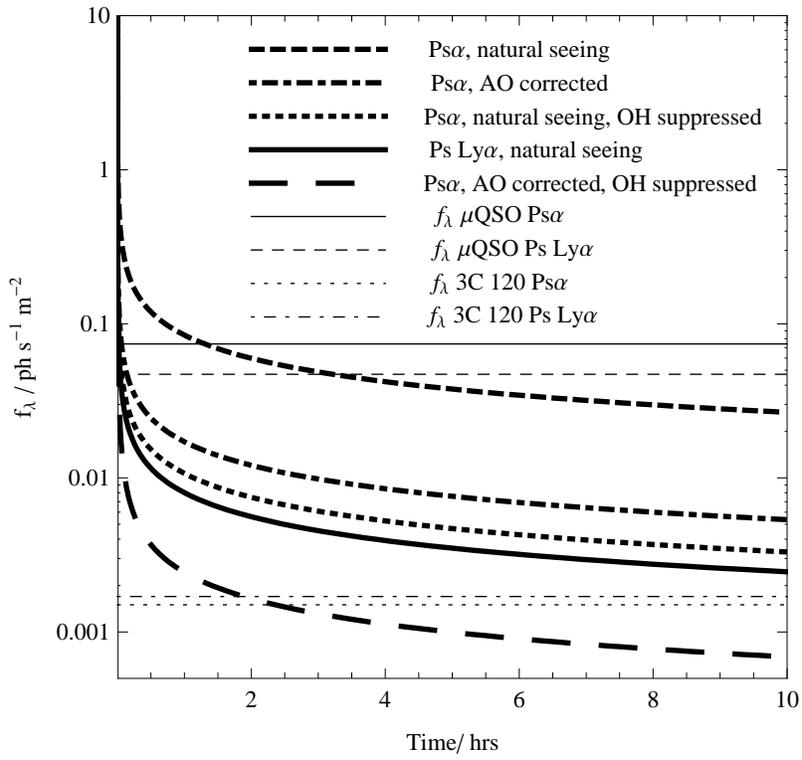}
\caption{The flux limits to obtain a signal to noise of 10 as a function of time for \psla\ and \psa\ under various assumptions of the background (see text for details).}
\label{fig:fluxlimits}
\end{figure}

In order to take into account all these various effects we have calculated the \psla\ and \psa\ flux limits to obtain a signal-to-noise of 10 as a function of time for backgrounds corresponding to \psla\ in natural seeing and \psa\ in natural seeing and adaptive optics corrected and with and without OH suppression.  The results are shown in Figure~\ref{fig:fluxlimits}.  (Note that here we quote the results as total line strengths since the signal to noise calculation requires integrating the emission line and the background over the line width.)  We assume that the Ps lines are resolved, and the backgrounds integrated over the line width are 0.0272, 13.05 and 0.197 ph s$^{-1}$ m$^{-2}$ arcsec$^{-2}$ for \psla, \psa\ and OH suppressed \psa\ respectively.  We assume that natural seeing is 1 arcsec in the UV and 0.5 arcsec in the NIR and that AO delivers a PSF of 0.1 arcsec FWHM.  We assumed a 8m telescope and a total system efficiency of 0.2.

The limiting fluxes are compared to the average expected flux for microquasars (specifically we compare to GROJ1655-40 the brightest confirmed mis-aligned microquasar, cf.\ Table~\ref{tab:mqso}) as calculated above in \S~\ref{sec:mqso}.  Both the \psla\ and the \psa\ lines should easily be detectable.    We also compare to the fluxes for 3C 120, assuming the same backgrounds as for Ps at $z=0$ (note that the NIR background is a strong function of wavelength so this assumption is approximate in the case of no OH suppression).  In this case the Ps emission lines are too faint to be detected in natural seeing and natural backgrounds for both \psla\ and \psa.   However with OH suppression and or AO correction \psa\ should be detectable.

\subsection{Simulations}

We have estimated the Ps recombination emission line strengths for a wide variety of sources; the expected fluxes are fainter than current observational limits for \psla\ and marginal for \psa.  However with the development of OH suppressing technologies  and the future launch of the JWST (\citealt{gar06}) we are on the threshold of a new era in NIR spectroscopy with much lower backgrounds and consequently much more sensitivity.  The improved sensitivity will make the detection of \psa\ lines feasible.  

The most promising targets for observation are mis-aligned microquasars and active galactic nuclei.  Emission from other candidates is generally too diffuse.  Mis-aligned microquasars should have a spot of Ps emission where the jet impacts the secondary star.  AGN should have similar spots of emission where the jet impacts gas clouds surrounding the nucleus.  In the case of AGN there will be bright emission from the nucleus itself, which must be subtracted off to reveal the faint Ps emission.

\begin{figure}
\centering \includegraphics[scale=1.0]{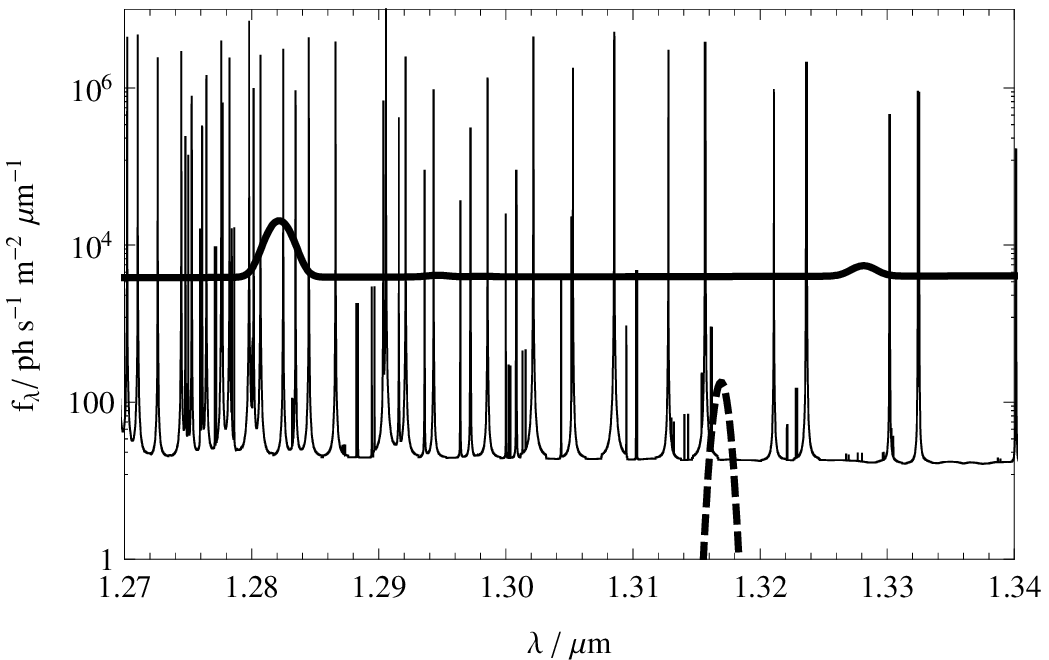}
\caption{Comparison of the components of the simulated spectrum of \psa\ emission from an AGN.  The emission from the AGN is shownby the thick black line, and is based on observations of NGC~4151 by \protect\citet{sto09}.  The \psa\ emission line is shown by the dashed black line and is based on calculations of the \pos\ production of 3C120 by \protect\citet{mar07}.  The near-infrared background is shown by the thin black line and is based on the models of \protect\citet{ell08}.}
\label{fig:ngc4151sim}
\end{figure}

We have simulated observations of \psa\ from an AGN.  We use the observations of NGC~4151 by \citet{sto09} as our template for the the AGN emission.  Specifically we model the spectrum of the region 0.7 arcsec E of the nuclues (see their figure 2, panel B), taking emission line fluxes from their table 1 and using a continuum value of $6 \times 10^{-16}$ erg s$^{-1}$ cm$^{-2}$ \AA$^{-1}$.  We model the \psa\ line by taking the flux for 3C120 based on the calculations of \citet{mar07} (see Table~\ref{tab:source_summary}) and scaling it to the redshift of NGC~4151.  The background spectrum is taken from \citet{ell08}.  The simulated components of the spectrum are shown in Figure~\ref{fig:ngc4151sim}.

The simulations assume OH suppressed observations on an 8m telescope with a 1 arcsec diameter fibre bundle as described in \citet{ell08}, with adaptive optics correction delivering a Strehl ratio of 0.3.  The spectral resolution was assumed to be  $R=3000$.  The simulations assume a varying night-sky  background and systematic errors in wavelength calibrations between reads.  For further details see \citet{ell08}.

Since the AGN continuum is much brighter than the \psa\ it must be treated as background, and observations will be required of the AGN plus \psa\ and of the AGN alone.  For this reason we anticipate that integral field spectroscopy will allow the greatest chance of a successful detection, since fluxes from many regions of the AGN can be  differenced, nullifying the requirement to know \emph{a priori} in which region Ps might be forming.

\begin{figure}
\centering \includegraphics[scale=0.6]{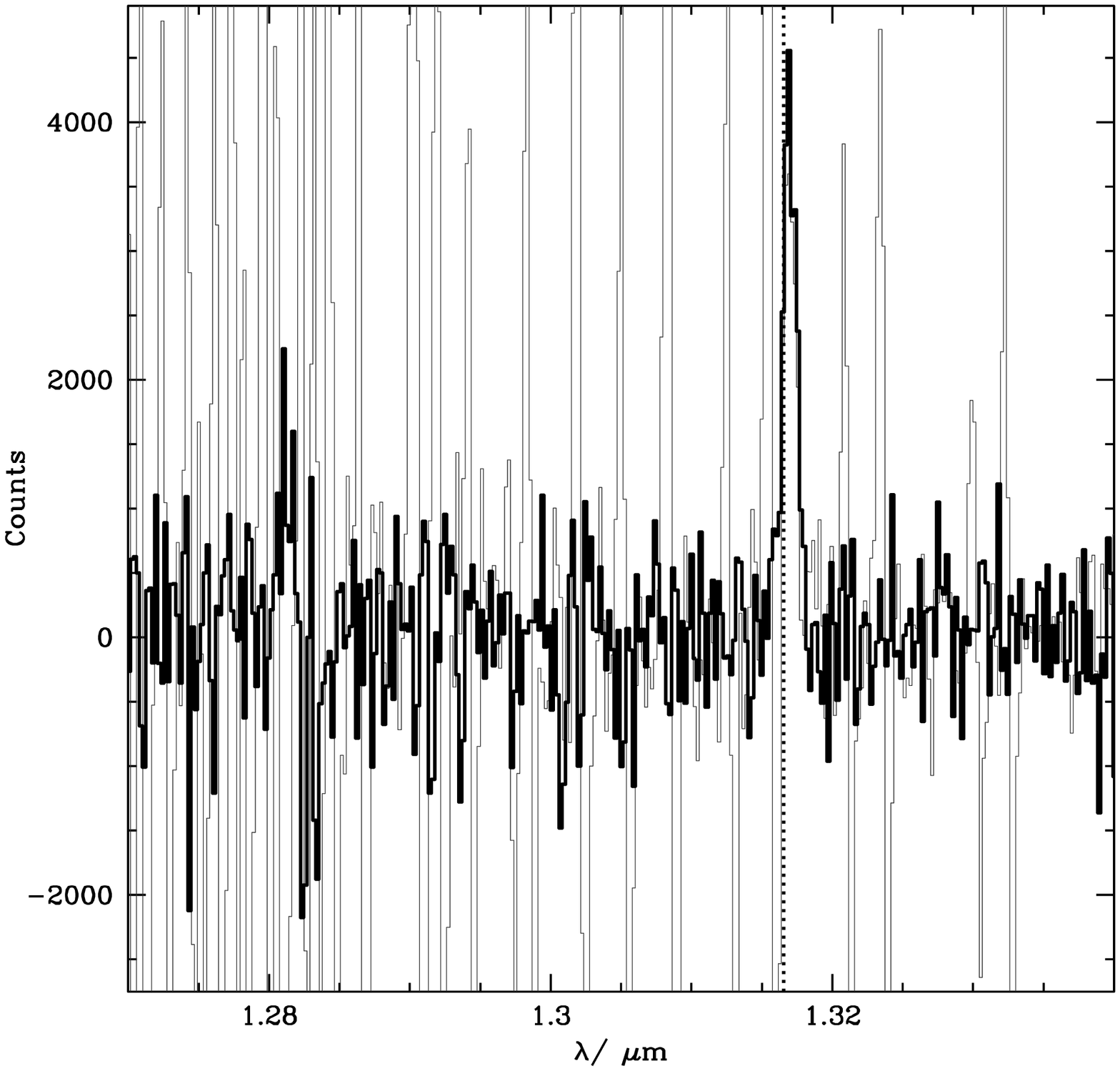}
\caption{A background and continuum subtracted simulated observation of \psa\ from an AGN.   The simulations are shown with OH suppression by the thick black line, and without by the thin grey line.  The dashed line shows the expected location of the \psa\ line.  The line is clearly detected in the OH suppressed observation.}
\label{fig:ngc4151psa}
\end{figure}

For the purposes of the simulation we assume that the AGN continuum can be precisely scaled between these two locations (i.e.\ in the simulations we assume the AGN emission does not change).  The simulation assumes a 6hr observation (composed of 12$\times$30 min individual exposures) of both regions.  The resulting background and continuum subtracted spectrum is shown in Figure~\ref{fig:ngc4151psa}.  The \psa\ line is clearly visible.  Without OH suppression the \psa\ is much fainter than the noise from the residual sky lines, and any detections would be marginal.

\section{SUMMARY AND DISCUSSION}
\label{sec:summ}

The existence of astrophysical sources of Ps is revealed through observations of Ps annihilation radiation emanating from the Galactic bulge, with a weaker component originating in the disc (e.g.\ \citealt{wei08b}).  It is expected that those Ps atoms in the triplet state will emit recombination emission lines prior to annihilation (\citealt{mcc84}).  The wavelengths of the recombination spectral lines will be twice those of the corresponding hydrogen lines (\citealt{moh34}).  In this paper we have assessed the possibility of observing the Ps recombination spectrum.

In \S~\ref{sec:phys} we reviewed the physics of the formation, annihilation and recombination spectrum of Ps.   It was argued that Ps formation will not take place in most astrophysical environments until the \pos\ are in thermal equilibrium with the ISM.  The reason is that \pos\ are produced in high energy processes, and they are initially relativistic.  They must therefore lose energy before they can radiatively recombine with a free electron or undergo charge exchange with a H atom to form Ps.  We note that for photoionised astrophysical environments with temperatures in the range $10^{3}$ -- $10^{6}$ K, Ps formation by radiative recombination dominates both charge exchange and  the direct annihilation of \pos\ with bound or free electrons.  This will be this case in close proximity to most astrophysical sources of Ps, which will also emit strong ultraviolet radiation which will ionise the surrounding medium.  We emphasise that this is not the case for most \pos, which will travel far from their origin (\S~\ref{sec:diffuse}); in a collisionally ionised medium charge exchange dominates radiative recombination for temperatures $\lsim 10^{5}$ K (\citealt{mcc84}).   However, in the case of point sources of Ps recombination emission lines, we are interested in the regime in which the Ps is formed in close proximity to the source and therefore assume that radiative recombination dominates.  Following these arguments, we gave formul\ae\ to calculate the Ps emission line strengths for sources of known \pos\ production rate under different conditions in the ISM.  We also developed a formula to convert  511keV fluxes to Ps recombination line fluxes.

In \S~\ref{sec:sources} we reviewed possible sources of \pos, both Galactic and extragalactic, and estimated \pos\ production rates from the literature.  The \pos\ production rates were used to calculate Ps emission line strengths.  Taking into account \pos\ diffusion in the ISM (cf.\ \S~\ref{sec:diffuse} and \citealt{jea06}) showed that for sources in which the \pos\ are ejected isotropically into the ISM the resulting Ps emission line has very low surface brightness.  Thus whilst sources such as SNe, nov\ae\ and pulsars may produce a potentially significant fraction of the total Galactic \pos\ budget, they are not promising candidates for Ps recombination line observations.

On the other hand sources such as microquasars, LMXBs and AGN jets are potentially promising candidates in which to observe Ps recombination lines.  The reason being that the \pos\ are anisotropic and may therefore produce a point source of emission if the jet collides with a dense cloud in the ISM.  For example in mis-aligned microquasars the jet impacts the secondary star, in LMXBs the jet may collide with stellar winds, and in AGN the jet will impact the gas clouds surrounding the galactic nucleus.  

In \S~\ref{sec:expt} we examined the prospects for detecting such sources.  We first compared our predictions of Ps recombination line fluxes to existing observational constraints.  All predicted fluxes are lower than current upper limits to the Ps emission line fluxes.  We therefore examined the sensitivity of current instrumentation, cf. Figure~\ref{fig:fluxlimits}.    The brightest mis-aligned microquasars should be detectable with current technology; more moderate sources are below current sensitivity limits.  The brightest AGN are below the sensitivity of current instrumentation.  In order to observe large samples of objects, e.g.\ to constrain the Galactic positron budget for a particular class of sources, requires deeper observations than are currently feasible.

The flux limits of observations of \psla\ are rather sensitive to the extinction, sources with low extinction will be brighter at \psla\ than \psa\ (cf.\ Figures~\ref{fig:extinct_total} and ~\ref{fig:extinct_density}).   Thus deep observations with careful continuum subtraction may yield \psla\ detections.

The current handicap to observations of \psa\ is the brightness of the background; this situation will not last for much longer.
Several imminent advances in near-infrared spectroscopy will allow much deeper observations than currently possible.  In particular the long standing problem of the night-sky is very close to being solved both by the launch of the JWST, and through photonic OH suppression from the ground (\citealt{bland04,bland08}).  Combined with advances in adaptive optics and NIR detector technology  the prospect of detecting \psa\ in the near future is propitious, cf.\ Figure~\ref{fig:fluxlimits}.

The successful detection of Ps recombination lines has the potential to yield immediate advances in our knowledge of several topics of interest.  Observations of Ps recombination lines at the location of AGN jet-cloud interactions would provide direct confirmation that jets are composed of a pair plasma -- a long standing problem which is still controversial (e.g.\ \citealt{cel93}; \citealt{rey96}; \citealt{war98}; \citealt{hir05}).  We suggest an observing campaign of several classes of AGN, e.g.\ targeting the regions around the jets of blazars, offering a potentially unobscured view of the most energetic region of the AGN, and the possibility of Ps formation in the regions adjacent to the jet.  Similarly observations of the radio lobes of AGN offer the possibility of detecting the location at which the \pos\ surviving the jet thermalise with the ISM.

An immediate and direct advance in our knowledge of the origin Galactic \pos\ would result from a detection of Ps emission from a class of Galactic sources such as microquasars.  This would place direct constraints on the significance of such sources the Galactic \pos\ budget, and indirect constraints on the significance of other sources through comparison with the total \pos\ budget from 511keV observations.
Similarly,  such detections would place constraints on the astrophysical processes responsible for the production of \pos\ in the sources.

Looking further into the future, one can envisage more ambitious experiments.  For example,
many models of dark matter invoke a super-symmetric dark matter particle.  In this case the dark matter particle will annihilate (albeit with a small cross section), resulting in the eventual production of an \el-\pos\ pair (see e.g.\ \citealt{boe04}).  Such a mechanism has been proposed to account for the Galactic 511keV radiation described above.  If significant constraints can be placed on the astrophysical origin of Galactic \pos\ this may have implications on the masses or existence of super-symmetric dark matter particles.
If dark matter were found to be a significant source of Galactic \pos\ then one could in principle use observations of Ps recombination lines to infer the masses and density profiles of dark matter haloes.  Such experiments are not feasible with any envisaged telescopes, but are a theoretically interesting possibility.

\section{ACKNOWLEDGMENTS}
We thank the referee for useful comments which have improved this manuscript.
We thank Alan Marscher for advice on the calculation of the  \pos\ flux in AGN jets.  JBH is supported by a Federation Fellowship from the Australian
Research Council.

\bibliographystyle{scemnras}
\bibliography{ps}

\end{document}